\documentclass[aps,prx,preprint,amsmath,amssymb,floatfix,citeautoscript]{revtex4-1}

\usepackage[T1]{fontenc}
\usepackage[utf8]{inputenc}
\usepackage{verbatim}
\usepackage{amsmath, amssymb, amsfonts, bm, amsthm, braket}
\usepackage{graphicx}
\usepackage{color}
\usepackage{hyperref} 

\DeclareMathOperator{\Tr}{Tr}

\newcommand{\xnotk}[1]{x_{0_k}^{(#1)}}
\newcommand{\xbark}[2]{\bar{x}_{0_k}^{#1 #2}}
\newcommand{\deltaxk}[2]{\Delta{x}_{0_k}^{#1 #2}}

\newcommand{\Gg}{\mathcal{G}}
\newcommand{\Hh}{\mathcal{H}}

\begin{document}
\date{\today}
\flushbottom

\title{Perturbation theory under the truncated Wigner approximation reveals how system-environment entanglement formation drives quantum decoherence}
\author{Justin Provazza\footnote{justin.provazza@northwestern.edu}}
\affiliation{Department of Chemistry, Northwestern University, Evanston, Illinois 60208, USA}
\author{Roel Tempelaar\footnote{roel.tempelaar@northwestern.edu}}
\affiliation{Department of Chemistry, Northwestern University, Evanston, Illinois 60208, USA}

\begin{abstract}
Quantum decoherence is the disappearance of simple phase relations within a discrete quantum system as a result of interactions with an environment. For many applications, the question is not necessarily how to avoid (inevitable) system-environment interactions, but rather how to design environments that optimally preserve a system's phase relations in spite of such interactions. The formation of system-environment entanglement is a major driving mechanism for decoherence, and a detailed understanding of this process could inform strategies for conserving coherence optimally. This requires scalable, flexible, and systematically improvable quantum dynamical methods that retain detailed information about the entanglement properties of the environment, yet very few current methods offer this combination of features. Here, we address this need by introducing a theoretical framework wherein we combine the truncated Wigner approximation with standard time-dependent perturbation theory allowing for computing expectation values of operators in the \textit{combined} system-environment Hilbert space. We demonstrate the utility of this framework by applying it to the spin-boson model, representative of qubits and simple donor-acceptor systems. For this model, our framework provides an analytical description of perturbative contributions to expectation values. We monitor how quantum decoherence at zero temperature is accompanied by entanglement formation with individual environmental degrees of freedom. Based on this entanglement behavior, we find that the selective suppression of low-frequency environmental modes is particularly effective for mitigating quantum decoherence.
\end{abstract}

\maketitle

\section{Introduction}
\label{sec:intro}
Discrete quantum systems serve as elementary building blocks of devices for quantum information,\cite{lauk2020perspectives} quantum sensing,\cite{degen2017quantum} and optoelectronic technologies.\cite{cao2020quantum,cheng2009dynamics} The inevitable presence of environmental interactions leads to a loss of coherence; simple phase relations between system states. Understanding this process is crucial for applications ranging from quantum computing\cite{nielsen2000quantum,balasubramanian2009ultralong} to solar energy harvesting.\cite{scholes2017using,cao2020quantum} In the former case it would be desirable to approach the limit of indefinitely retained coherence.\cite{devoret2013superconducting} This limit is commonly envisioned as that of minimal environmental reorganization energy and minimal temperature, which, however, renders implementations of quantum computing devices impractical. A major driving mechanism of decoherence is the formation of system-environment entanglement, and it is conceivable that decoherence can be suppressed by rationally controlling how this entanglement evolves, although the number of studies exploring these principles has remained limited.\cite{eisert2002quantum,hilt2009system,pernice2011decoherence,rossatto2011purity,pernice2012system,roszak2015characterization,costa2016system,salamon2017entanglement,roszak2018criteria} This would suggest implementation strategies akin to those found in natural light harvesting, where discrete quantum systems in the form of coupled chromophores experience an environment tailored to their functioning: the efficient dynamical localization of quantum excitations on an acceptor state.\cite{dean2016vibronic, lee2017first}

Modeling the transient behavior of discrete quantum systems \textit{and} their environments can provide mechanistic insight on how details of an initial state preparation and system-environment interactions impact their evolution, \cite{dodin_provazza} while offering an opportunity to optimize function by engineering these factors. There has been an enormous effort in the development of efficient nonadiabatic quantum dynamics techniques that are capable of reliably modeling decoherence processes following preparation of an initial nonequilibrium state. In spite of this, however, there remains a lack of modeling techniques offering explicit insight into how individual environmental modes become entangled with the system states. Such explicit insight becomes lost in standard quantum master equation techniques such as Redfield theory,\cite{redfield1957theory} F\"orster theory, \cite{forster1949experimentelle} and the Lindblad master equation \cite{lindblad1976generators} as well as their subsequent extensions,\cite{scholes2001adapting,hwang2015coherent,trushechkin2019calculation,montoya2015extending,breuer2002theory} where a reduced density matrix of the system is obtained by ``projecting out'' the environmental degrees-of-freedom (DOFs). This adds to the inflexibility of such techniques with regard to a smallness parameter used in their perturbative expansions. Environmental details are similarly inaccessible in influence functional formalisms, such as the numerically exact hierarchical equations of motion\cite{tanimura1989time,tanimura2020numerically} (HEOM), which only capture how a system evolves under the influence of an environment. Semiclassical and phase space methods \cite{tully1971trajectory,miller2001semiclassical,polkovnikov2003quantum,polkovnikov2010phase,huo2011communication,ananth2007semiclassical,tully1990molecular} may provide an unreliable description of environmental entanglement by approximating the involved DOFs as being classical, which may additionally lead to spurious results\cite{lee2016semiclassical} when there are high frequency environmental modes involved.\cite{provazza2019multi} Arguably, the most versatile technique capable of describing system-environment entanglement in detail is multi-configuration time-dependent Hartree\cite{beck2000multiconfiguration,kuhn2016mlmctdh} which, however, suffers from high computational cost, especially with increasing number of environmental modes.

In this work, we present an approximate yet systematically-improvable framework for modeling decoherence of a discrete quantum system interacting with an environment at low computational cost while offering access to the entanglement formation of individual environmental DOFs. The framework is based on the truncated Wigner approximation (TWA) as derived through a phase space path integral analysis. We show for a ubiquitous harmonic environment model how under the TWA one can obtain perturbative expressions to arbitrary order through a quantum momentum fluctuation analysis,\cite{polkovnikov2003quantum,polkovnikov2010phase,AnalyticVibronicSpectra} even when the perturbative terms contain arbitrary order environmental operator dependence. This flexibility allows one to perform rotations of the system basis, while being fully capable of capturing the resulting environmental operator dependence of the associated Hamiltonian matrix elements. In contrast to common perturbative methods based on projection operator techniques, our framework also avoids the assumption that the initial density operator is of factorized form. Instead, it only imposes that the initial environmental Wigner function be of a Gaussian form that encompasses both the thermal density and (shifted) wave packets, which may be utilized as a means to describe configurations of a single discrete quantum system.\cite{dodin_provazza} For the harmonic environment model considered here, our framework provides a fully consistent avenue for deriving (nearly) analytical time-dependent expectation values of operators in the system \text{and} environment Hilbert space, including that of entanglement entropy.

To demonstrate the capabilities of our framework, we consider various versions of the spin-boson model, representative of qubits and simple donor-acceptor systems. Simulations are presented for a qubit interacting with a continuous harmonic environment at zero temperature, where the ability of our framework to compute expectation values in the environment Hilbert space is utilized by monitoring the time-dependent entanglement entropy of individual environmental modes. Through this, the entanglement formation accompanying quantum decoherence of the qubit is demonstrated to particularly involve low-frequency environmental DOFs. Directed by this result, we show that the selective suppression of low-frequency modes significantly mitigates quantum decoherence compared to the selective suppression of high-frequency modes for a fixed reorganization energy, showing that reorganization energy itself is an incomplete metric for an estimate of quantum decoherence. We further demonstrate the flexibility of our framework in treating perturbations regardless of their environmental operator dependence by considering a weakly interacting environment as well as an environment consisting of a single (discrete) mode. To this end, we derive perturbative reduced density matrix expressions where the Hamiltonian is represented in two contrasting system bases that lead to environmental operator independent and dependent perturbations. The importance of this flexibility is highlighted by considering parameters where the two perturbative expansions result in different accuracies at second order.

The present manuscript is outlined as follows: In Section \ref{sec:background} we introduce the perturbative expansion under the TWA that serves as the basis of our framework. Next, in Section \ref{sec:HarmonicModel} we present a simple harmonic environment model for which we derive expressions for $N$th order perturbative contributions to the expectation value of operators in the combined system-environment Hilbert space. In Section \ref{sec:QubitDissipation} we evaluate the entanglement formation for a qubit interacting with a continuous, harmonic environment, including the selective suppression of environmental DOFs. In Section \ref{sec:EnvironmentalDependence} we present results for the weakly interacting and single-mode environments. Finally, in Section \ref{sec:conclusions} we provide concluding remarks and comment on potential applications for the framework described here.

\section{Perturbation theory under the Truncated Wigner Approximation}
\label{sec:background}
We begin by considering a discrete quantum system in contact with an environment, described by (diabatic) basis states $\{\ket{n}\}$ and phase space operators $(\mathbf{\hat{x}},\mathbf{\hat{p}})$, respectively. The Hamiltonian can be partitioned as
\begin{equation}
    \hat{H}(t) = \hat{H}_0 + \hat{V}(t),
\end{equation}
with a reference Hamiltonian, $\hat{H}_0$, that is chosen to be diagonal in the system basis as
\begin{equation}\label{eq:ReferenceHamiltonian}
    \hat{H}_0 = \sum_{k}\frac{\hat{p}_k^2}{2} + \sum_{n}  U_{n}(\mathbf{\hat{x}}) \ket{n}\bra{n}.
\end{equation}
Here, $\hat{p}_k$ is the mass-weighted momentum operator of the $k$th environmental DOF. For now, we leave the system state-dependent potentials, $U_{n}(\mathbf{\hat{x}})$, arbitrary with the understanding that they are analytic functions of the environmental position operators, $\mathbf{\hat{x}}$. The remaining term in the Hamiltonian, which accounts for off-diagonal contributions in the system basis, is considered to be a perturbation of the form
\begin{equation}\label{eq:Perturbation}
    \hat{V}(t) = \sum_{m \ne n} V_{mn}(\mathbf{\hat{x}},t) \ket{m} \bra{n}.
\end{equation}
Here, $V_{mn}(\mathbf{\hat{x}},t)$ is a perturbation operator matrix element in the system basis and is, in general, a function of environmental position operators and time. With these definitions, all transitions between system states are dictated by the perturbative operation. \footnote{The specification of a purely off-diagonal perturbation is of course not necessary, and is simply chosen out of convenience for the analyses to follow.}

The time-dependent density operator can be expanded in a perturbative series as
\begin{equation}\label{eq:Expansion}
\hat\rho(t) = \sum_{N = 0}^\infty \hat\rho_N(t),
\end{equation}
where the $N$th order density operator, for $N\ge1$, is provided by
\begin{equation}\label{eq:NthOrder}
\begin{split}
& \hat{\rho}_N(t) = \left( \frac{-i}{\hbar}\right)^N \left\{ \prod_{j=1}^N \int_0^{\tau_{j+1}} d\tau_j \right\} e^{-\frac{i}{\hbar}\hat{H}_0 t} \left[ \hat{V}_I(\tau_N),[\hat{V}_I(\tau_{N-1}),\dots[\hat{V}_I(\tau_1),\hat{\rho}(0)]\dots]]\right]e^{\frac{i}{\hbar}\hat{H}_0 t}.
\end{split}
\end{equation}
Here, $\hat{V}_I(\tau_j) = e^{\frac{i}{\hbar}\hat{H}_0 \tau_j}\hat{V}(\tau_j)e^{-\frac{i}{\hbar}\hat{H}_0 \tau_j}$ is the perturbation operator in the interaction representation and $\tau_{N+1} = t$. The 0th order contribution is given by $\hat{\rho}_0(t) = e^{-\frac{i}{\hbar}\hat{H}_0 t} \hat{\rho}(0)e^{\frac{i}{\hbar}\hat{H}_0 t}$.

Given the definitions of the reference Hamiltonian and perturbation provided above, we can express the time-dependent expectation value of an operator as a sum over perturbative contributions,  $\langle\hat{O}(t)\rangle = \Tr[\hat{O}\hat{\rho}(t)] = \sum_{N=0}^{\infty} \langle \hat{O} (t)\rangle_N$. In defining this, we have simply used Eq. (\ref{eq:Expansion}) for the definition of the time-evolved density operator, so that $\langle \hat{O} (t)\rangle_N=\Tr[\hat{O}\hat{\rho}_N(t)]$. Since we have chosen the reference Hamiltonian to only contain contributions that are diagonal in the system basis, we can apply the truncated Wigner approximation (TWA) as described elsewhere\cite{Shi2004,McRobbie2009,AnalyticVibronicSpectra} along with a quantum momentum fluctuation analysis \cite{polkovnikov2003quantum,polkovnikov2010phase} (as outlined in the Supporting Information) to recover a generally-approximate functional form for the $N$th order contribution to the operator expectation value as
\begin{equation}\label{eq:TWAPerturbativeResult}
\begin{split}
\langle \hat{O}(t) \rangle_N \approx  \left( \frac{-i}{\hbar}\right)^N \left\{ \prod_{j=1}^N \int_0^{\tau_{j+1}}d\tau_j\right\} \sum_{\{n_j,n'_j\}} \ & \int \frac{d\mathbf{x}_{0}d\mathbf{p}_{0}}{(2\pi\hbar)^D} \ W_{n_0,n'_0}(\mathbf{x}_{0},\mathbf{p}_{0})   \\
 \times \bigg[ \Theta^{\{n_j,n'_j\}}_N\left(\left\{\big(\mathbf{\tilde{x}}_{\text{Bopp}}^{(j)}(\tau_j),\tau_j\big) \right\}\right)   & O_W^{n'_N,n_N}(\mathbf{x}^{(N)}_{\delta p}(t),\mathbf{p}^{(N)}_{\delta p}(t)) \\
 & \times e^{-\frac{i}{\hbar}\sum_{j=0}^N \int^{\tau_{j+1}}_{\tau_j} ds \left[U_{n_{j}}(\mathbf{x}^{(j)}_{\delta p}(s)) -U_{n'_{j}}(\mathbf{x}^{(j)}_{\delta p}(s)) \right]} \bigg]_{\delta p = 0},
\end{split}
\end{equation}
where we have taken $\tau_0=0$ and $\tau_{N+1} = t$ to represent the initial and final times, respectively. Here and throughout, forward (backward) propagator matrix elements are labeled with unprimed (primed) system state indices, and indices within curly brackets involve the complete set $j=0,1,\ldots,N$, unless noted otherwise. We further have that
\begin{equation}\label{eq:PartialWigner}
W_{n,n'}(\mathbf{x},\mathbf{p}) = \int d\mathbf{z} \bra{\mathbf{x} + \frac{\mathbf{z}}{2},n}\hat{\rho}_0 \ket{n',\mathbf{x} - \frac{\mathbf{z}}{2}} e^{-\frac{i}{\hbar}\mathbf{p}\cdot\mathbf{z}}
\end{equation}
is the system state-dependent Wigner function,\cite{wigner1932,huo2011communication} and $O_W^{n,n'}(\mathbf{x},\mathbf{p})$ is the system state-dependent Weyl symbol\cite{weyl1927quantenmechanik} of the operator $\hat{O}$ that is defined analogously to the Wigner function (with the operator, $\hat{O}$, in place of the density operator). 

In Eq. (\ref{eq:TWAPerturbativeResult}), we have introduced \textit{shifted} environmental trajectories, $\mathbf{x}^{(j)}_{\delta p}(s)$. These shifted trajectories describe ``classical'' evolution supplemented with infinitesimal momentum fluctuations that accumulate each time a perturbation operator acts on (and changes the state of) the system. Their time-evolution during the $j$th time segment, between the $j$th and $(j+1)$th perturbative interaction, is obtained by integrating Hamilton's equations under the influence of an average quantum force as \footnote{The force defining the classical time evolution should generally be derived through the Weyl symbol of the Hamiltonian\cite{polkovnikov2010phase}. However, since there are no mixed position-momentum terms contributing to the state-dependent potential, the Weyl symbol of the Hamiltonian is simply obtained by replacing the quantum operators with classical numbers.}
\begin{equation}\label{eq:InstantaneousVelocity}
\dot{\mathbf{x}}^{(j)}_{\delta p}(s) = {\mathbf{p}^{(j)}_{\delta p}(s)}
\end{equation}
\begin{equation}\label{eq:InstantaneousForce}
{\dot{\mathbf{p}}}^{(j)}_{\delta p}(s) = -\frac{1}{2} \nabla_{{\mathbf{x}}_{\delta p}}\left(U_{n_j}({\mathbf{x}}^{(j)}_{\delta p}(s)) + U_{n'_j}(\mathbf{x}^{(j)}_{\delta p}(s)) \right).
\end{equation}
Here, the initial point in phase space for the $j$th evolution segment is provided by the endpoint of the previous $(j-1)$th segment supplemented with an additional infinitesimal momentum fluctuation, $\mathbf{\delta p}_{\tau_j}$, as $(\mathbf{x}^{(j)}_{\delta p}(\tau_j),\mathbf{p}^{(j)}_{\delta p}(\tau_j)) = (\mathbf{x}^{(j-1)}_{\delta p}(\tau_j),\mathbf{p}^{(j-1)}_{\delta p}(\tau_j) + \mathbf{\delta p}_{\tau_j})$. However, phase space initial conditions for the first time segment are distributed according to the partial Wigner transform of the initial density operator as defined in Eq. (\ref{eq:PartialWigner}), and they do not contain any explicit momentum fluctuation dependence. 

Each Liouville space pathway (sequence of system state transitions) that contributes to Eq. (\ref{eq:TWAPerturbativeResult}) through a series of perturbative interactions is weighted proportionally to a tensor element, $\Theta^{\{n_j,n'_j\}}_N\left(\left\{\big( \mathbf{\tilde{x}}_{\text{Bopp}}^{(j)}(\tau_j),\tau_j\big) \right\}\right)$. In general, these weights take on a different value for each contribution to the nested sum over system states in Eq. (\ref{eq:TWAPerturbativeResult}). Each element, in general, depends both on the times at which each perturbation operator acts as well as time-dependent Bopp operators defined as\cite{bopp1961statistische,polkovnikov2010phase} 
\begin{equation}\label{eq:BoppOpp}
 \mathbf{\tilde{x}}_{\text{Bopp}}^{(j)}(\tau_j) \equiv \mathbf{x}^{(j-1)}_{\delta p}(\tau_j) - S_j\frac{i\hbar}{2}\frac{\partial}{\partial\mathbf{\delta p}_{\tau_j}}.
\end{equation}
The emergence of Bopp operators within this perturbative framework is necessary in order to completely describe the impact of the perturbation's environmental operator dependence, as shown in the Supporting Information. This analysis reveals that one must evaluate the phase factors and Weyl symbol in Eq. (\ref{eq:TWAPerturbativeResult}) along shifted trajectories as defined in Eqs. (\ref{eq:InstantaneousVelocity}) and (\ref{eq:InstantaneousForce}), while the perturbation operator at time $\tau_j$ is evaluated along time-dependent Bopp operators as
\begin{equation}\label{eq:PerturbationMatrixElementBopp}
    V_{mn}(\mathbf{\hat{x}},\tau_{j}) \rightarrow V_{mn}\left(\mathbf{\tilde{x}}_{\text{Bopp}}^{(j)}(\tau_j),\tau_{j}\right).
\end{equation}
In this expression, the Bopp operators given in Eq. (\ref{eq:BoppOpp}) are to be evaluated at $S_j=+1$ when the perturbation operates on the left (ket) and at $S_j=-1$ when it operates on the right (bra). 

In the implementation of Eq. (\ref{eq:TWAPerturbativeResult}), the derivatives in the definition of the Bopp operator are allowed to operate on \textit{everything} containing dependence on the corresponding momentum fluctuation. This, in general, consists of the phase factor and Weyl symbol as well as any later-time perturbative operator's dependence on the shifted environmental trajectory. When the perturbation is chosen to be solely off-diagonal in the system basis, a tensor element is of the simple form
\begin{equation}\label{eq:ThetaGeneral}
\begin{split}
& \Theta^{\{n_j,n'_j\}}_N\left(\left\{\big(\mathbf{\tilde{x}}_{\text{Bopp}}^{(j)}(\tau_j),\tau_j\big) \right\}\right) \\
& = \prod_{j=1}^{N} \bigg( V_{n_{j}n_{j-1}}(\mathbf{x}^{(j-1)}_{\delta p}(\tau_j) - \frac{i\hbar}{2}\frac{\partial}{\partial\mathbf{\delta p}_{\tau_j}},\tau_{j}) \delta_{n'_{j-1}n'_{j}}-V_{n'_{j-1}n'_{j}}(\mathbf{x}^{(j-1)}_{\delta p}(\tau_j) +\frac{i\hbar}{2}\frac{\partial}{\partial\mathbf{\delta p}_{\tau_j}},\tau_{j})  \delta_{n_{j}n_{j-1}}\bigg).
\end{split}
\end{equation}
The result is then evaluated at $\mathbf{\delta p}_{\tau_j}=0$ for all $\tau_j$.

For general potentials and Wigner functions, Eq. (\ref{eq:TWAPerturbativeResult}) clearly has very limited utility owing to an exponential scaling with respect to the perturbative order provided by the nested time integrals. In fact, it would be significantly less efficient than simply mapping the system projection operators onto bosonic raising and lowering operators and computing the appropriate matrix elements as prescribed by the so-called Meyer-Miller-Stock-Thoss mapping procedure.\cite{Stock1997,Meyer1979} The resulting Hamiltonian is then amenable to semiclassical algorithms, where the system and its environment are treated on an equal footing (\textit{i.e.}, through an ensemble of classical trajectories with quantum mechanically-distributed initial conditions).\cite{Stock1997,huo2011communication,ananth2007semiclassical,Shi2004} However, any nonzero coupling between the system and its environment is at least cubic in phase space operators and therefore not treated exactly within a semiclassical framework. Because of this, the accuracy of such semiclassical treatments relies on an inherent assumption that the system-environment coupling is weak or that the coupled DOFs are sufficiently ``classical''.\cite{polkovnikov2010phase,provazza2020modeling} While one may potentially describe quantum corrections to such approaches through a quantum fluctuation analysis similar to that described here,\cite{polkovnikov2010phase} the associated numerical convergence for large systems including hundreds (or more) environmental DOFs has proven challenging in our preliminary studies. The benefit of approaching the time evolution within a perturbative framework becomes apparent when one is interested in solving problems with system state-dependent potentials, $U_n(\mathbf{\hat{x}})$, that foster analytical expressions for the classical time-evolution of the environmental DOFs. This significantly simplifies the problem of solving Eq. (\ref{eq:TWAPerturbativeResult}) and yields analytical insight into time-dependent observables of quantum systems and their environments. \cite{AnalyticVibronicSpectra} The expression in Eq. (\ref{eq:TWAPerturbativeResult}) thus serves as the starting point for subsequent analyses in this manuscript and, in what follows, we utilize it to derive perturbative contributions to expectation values for a discrete quantum system coupled to a harmonic environment.

\section{Harmonic environment model}
\label{sec:HarmonicModel}
\begin{figure}[htb]
\includegraphics[width=0.5\linewidth]{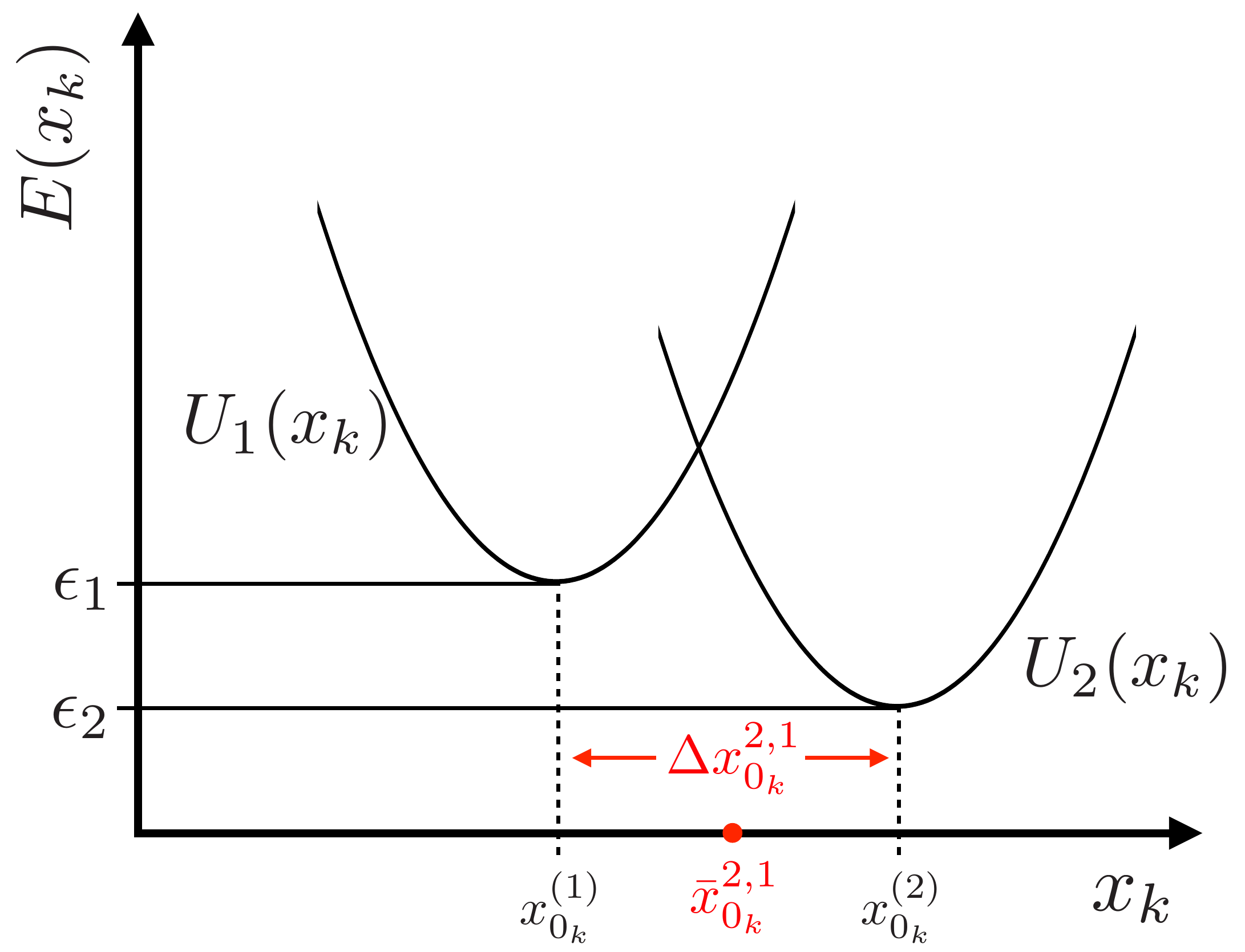}
\caption{Two potential energy surfaces contributing to the dynamics of the $k$th environmental mode given by $U_j = {\epsilon}_j + \frac{1}{2} \omega_k^2 (x_k - x_{0_k}^{(j)})^2$. Key quantities entering the expressions in the framework described in this manuscript are the mode frequency $\omega_k$, potential minimum $x_{0_k}^{(j)}$, the mean potential minimum $\xbark{j}{l} = \frac{1}{2}(x_{0_k}^{(j)}+x_{0_k}^{(l)})$, the potential minimum difference $\deltaxk{j}{l} = x_{0_k}^{(j)}-x_{0_k}^{(l)}$, and the energy of system state-dependent potential when evaluated at the potential minimum ${\epsilon}_j$.}
\label{fig:Potentials}
\end{figure}

For the harmonic environment model considered here, we define the system state-dependent potential as
\begin{equation}\label{eq:HarmonicModel}
U_n(\mathbf{\hat{x}}) = \epsilon_n + \frac{1}{2}\sum_{k} \omega_k^2 \left( \hat{x}_k - x_{0_k}^{(n)}\right)^2 ,
\end{equation}
where $\epsilon_n$ is the energy of the $n$th system state evaluated at its potential energy minimum, $\omega_k$ is the frequency of the $k$th environmental mode, and $x_{0_k}^{(n)}$ is the potential energy minimum for the $k$th mode when in the $n$th system state. This bi-linearly coupled harmonic model is ubiquitous in condensed matter and chemical physics, with the shifted potentials often representing the interaction of a molecular normal mode (or phonon) with, for example, electronic state $n$. When the system state-dependent potential is of the form above, where any pair of harmonic potentials differ \textit{only} by a relative shift in their minima (\textit{i.e.}, their frequencies are independent of the system state), the time-evolution of environmental DOFs as obtained through the TWA is given analytically and \textit{exactly} as a trajectory in phase space evolving on the mean potential of the active (forward and backward) system states.\cite{AnalyticVibronicSpectra,geva2021electronic,Shi2004,McRobbie2009}  In Figure \ref{fig:Potentials}, we provide a pictorial representation of two system state-dependent environmental potentials and highlight the key parameters that contribute to expressions in this our framework.

For a given sequence of system state transitions (or Liouville pathway), the time dependent position of the $k$th environmental DOF during the $j$th time segment (\textit{i.e.}, the evolution following $j$ perturbative interactions) is found to be
\begin{equation}\label{eq:ShiftedTrajectory}
x^{(j)}_{\delta p_k}(s) = x^{(j)}_{cl_k}(s) + \frac{1}{\omega_k}\sum_{l=1}^{j}\delta p_{{\tau_l}}^{(k)} \sin(\omega_k(s-\tau_l)).
\end{equation}
Here, the classical evolution following $j$ system state transitions is given as
\begin{equation}\label{eq:HarmonicModelTrajectories}
\begin{split}
x^{(j)}_{{cl}_k}(s)=x_{0_k}&\cos(\omega_k (s-\tau_0)) + \frac{p_{0_k}}{\omega_k}\sin(\omega_k (s-\tau_0))  + \bar{x}_{0_k}^{n_j n'_j}[1-\cos(\omega_k(s - \tau_j)) ]\\
& + \sum_{l=0}^{j-1} \bar{x}_{0_k}^{n_{l}n'_{l}}[\cos(\omega_k(s-\tau_{l+1}))-\cos(\omega_k( s -\tau_l)) ],\\
\end{split}
\end{equation}
where $s \ge \tau_j$, $\bar{x}_{0_k}^{n_j n'_j} = \frac{1}{2} ({x}_{0_k}^{(n_j)} + {x}_{0_k}^{(n'_j)})$ is the mean potential minimum between system states $n_j$ and $n'_j$, and the momentum is given by $p^{(j)}_{{cl_k}}(s)=\dot{x}^{(j)}_{{cl}_k}(s)$. We remind the reader that state labels without (with) a prime indicate states that are visited along the forward (backward) evolution of the system. For this shifted harmonic oscillator model, the evolution expressions for the environmental DOFs are obtained exactly within the TWA, irrespective of the perturbation operator's dependence on environmental operators.

With analytical expressions for the time-evolution of the environmental positions in-hand, along with the definitions of the system state-dependent potentials, one can readily evaluate the phase factors in Eq. (\ref{eq:TWAPerturbativeResult}) as
\begin{equation}\label{eq:HarmonicModelPhaseFactors}
\begin{split}
-\frac{i}{\hbar}\sum_{j=0}^N &\int^{\tau_{j+1}}_{\tau_j}  ds \left[U_{n_{j}}(\mathbf{x}^{(j)}_{\delta p}(s)) -U_{n'_{j}}(\mathbf{x}^{(j)}_{\delta p}(s)) \right] \\
 & = -\frac{i}{\hbar}  \sum_{j=0}^N \left[(\tilde\epsilon_{n_j} - \tilde\epsilon_{n'_j})(\tau_{j+1} - \tau_j)   - \sum_k \omega_k^2 \deltaxk{n_j}{n'_j} \int_{\tau_{j}}^{\tau_{j +1}} ds  \ x^{(j)}_{{cl}_k}(s)\right] \\
 & \hspace{1cm}  - \frac{i}{\hbar}  \sum_k\sum_{j=1}^{N} \delta p^{(k)}_{\tau_j} \sum_{l=j}^{N}  \left\{ \deltaxk{n_l}{n'_l} [ \cos(\omega_k(\tau_{l+1}-\tau_j)) - \cos(\omega_k(\tau_{l}-\tau_{j})) ]\right\},
\end{split}
\end{equation}
where $\tilde\epsilon_{n_j} = {\epsilon}_{n_j} + \frac{1}{2}\sum_{k} \omega_k^2x_{0_k}^{(n_j)2}$ is the energy of system state $n_j$ evaluated at the origin (\textit{i.e.}, $U_{n_j}(\mathbf{{x}})|_{\mathbf{{x}}=0}$), and $\deltaxk{n_j}{n'_j} = {x}_{0_k}^{(n_j)} - {x}_{0_k}^{(n'_j)}$ is the difference between potential minima that the $k$th environmental mode experiences when states $n_j$ and $n'_j$ are occupied. We readily find that
\begin{equation}
\begin{split}
&\int_{\tau_j}^{\tau_{j+1}} ds \  x^{(j)}_{cl_k}(s) = \frac{x_{0_k}}{\omega_k} [\sin(\omega_k \tau_{j+1}) - \sin(\omega_k \tau_j)] - \frac{p_{0_k}}{\omega_k^2}[\cos(\omega_k\tau_{j+1}) - \cos(\omega_k\tau_j)]  \\
& + \sum_{l=0}^{j-1} \frac{\bar{x}_{0_k}^{n_{l}n'_{l}}}{\omega_k}[\sin(\omega_k (\tau_{j+1}-\tau_{l+1})) - \sin(\omega_k (\tau_{j+1}-\tau_{l})) + \sin(\omega_k (\tau_{j}-\tau_l)) - \sin(\omega_k (\tau_j-\tau_{l+1}))]\\
& - \frac{\bar{x}_{0_k}^{n_jn'_j}}{\omega_k}\sin(\omega_k(\tau_{j+1} - \tau_j)) + \bar{x}_{0_k}^{n_jn'_j}(\tau_{j+1} - \tau_j),
\end{split}
\end{equation}
where we have used that $\tau_0=0$. Breaking the phase factors down into parts depending on momentum fluctuations and on environmental correlation functions, Eq. (\ref{eq:TWAPerturbativeResult}) is reformulated as
\begin{equation}\label{eq:TWAPerturbativeResultHarmonic1}
\begin{split}
\langle \hat{O}(t) \rangle_N = \left( \frac{-i}{\hbar}\right)^N \left\{ \prod_{j=1}^N \int_0^{\tau_{j+1}}d\tau_j\right\}  \sum_{\{n_j,n'_j\}} \  & e^{-\frac{i}{\hbar}\sum_{j=0}^N ({\epsilon}_{n_j}-{\epsilon}_{n'_j}) (\tau_{j+1} - \tau_j)} e^{ \Phi^{(N,\Hh)}_{\{n_j,n'_j\}}(\{\tau_j\}_{j=0}^{N+1})} \\
\times   \int \frac{d\mathbf{x}_{0}d\mathbf{p}_{0}}{(2\pi\hbar)^D} \  & e^{\frac{i}{\hbar} \sum_{k}  \left( \sum_{j=0}^N\deltaxk{n_j}{n'_j}[\sin(\omega_k \tau_{j+1}) - \sin(\omega_k \tau_j)]\right)\omega_k x_{0_k} } \\
\times \ & e^{- \frac{i}{\hbar} \sum_{k}  \left( \sum_{j=0}^N\deltaxk{n_j}{n'_j}[\cos(\omega_k \tau_{j+1}) - \cos(\omega_k \tau_j)]\right) p_{0_k} }  \\
\times \bigg[ \Theta^{\{n_j,n'_j\}}_N\left(\left\{\big(\mathbf{\tilde{x}}_{\text{Bopp}}^{(j)}(\tau_j),\tau_j\big) \right\}\right) \ O_W^{n'_N,n_N}&(\mathbf{x}^{(N)}_{\delta p}(t),\mathbf{p}^{(N)}_{\delta p}(t)) \ e^{ - \frac{i}{\hbar} \phi_N(\mathbf{\delta p}) }\bigg]_{\delta p = 0} W_{n_0,n'_0}(\mathbf{x}_{0},\mathbf{p}_{0}).
\end{split}
\end{equation}
The momentum fluctuation-dependent phase factor is given by
\begin{equation}\label{eq:MomentumFluctuationPhase}
 \phi_N(\mathbf{\delta p}) = \sum_k\sum_{j=1}^{N} \delta p^{(k)}_{\tau_j} \sum_{l=j}^{N}  \left\{ \deltaxk{n_l}{n'_l} [ \cos(\omega_k(\tau_{l+1}-\tau_j)) - \cos(\omega_k(\tau_{l}-\tau_{j})) ]\right\},
\end{equation}
while the influence of the harmonic environment is partially encoded through the combination of environmental correlation functions through
\begin{equation}\label{eq:AllHCorrelationFunctions}
\begin{split}
& \Phi^{(N,\Hh)}_{\{n_j,n'_j\}}(\{\tau_j\}_{j=0}^{N+1}) =  - \frac{i}{\hbar}\sum_{j=0}^N \mathcal{H}_{n_jn'_j}^{n_jn'_j}(\tau_{j+1} - \tau_j) \\ 
& + \frac{i}{\hbar} \sum_{j=1}^N\sum_{l=0}^{j-1}\left\{ \mathcal{H}_{n_ln'_l}^{n_jn'_j}(\tau_{j+1} - \tau_{l+1}) - \mathcal{H}_{n_ln'_l}^{n_jn'_j}(\tau_{j+1} - \tau_l) +\mathcal{H}_{n_ln'_l}^{n_jn'_j}(\tau_{j} - \tau_l) - \mathcal{H}_{n_ln'_l}^{n_jn'_j}(\tau_{j} - \tau_{l+1})\right\}.
\end{split}
\end{equation}
Here, the environmental correlation functions are given by
\begin{equation}\label{eq:HCorrelationFunction}
\mathcal{H}^{ab}_{cd}(t)= \frac{1}{\pi} \int_0^\infty d\omega \ \frac{1}{\omega^2}\left(J_{ac}(\omega) + J_{ad}(\omega)-J_{bc}(\omega) -J_{bd}(\omega)\right) \sin(\omega t),
\end{equation}
with spectral densities (summarizing the interaction of a system with a harmonic environment) defined in terms of the frequency and contributing potential minima of each environmental mode as 
\begin{equation}\label{eq:SpectralDensityDefinition}
J_{ab}(\omega) = \frac{\pi}{2}\sum_k \omega_k^3 x_{0_k}^{(a)}x_{0_k}^{(b)}\delta(\omega-\omega_k).
\end{equation}
We note that, with the definition of the correlation function in Eq. (\ref{eq:HCorrelationFunction}), we get no contribution when $a=b$, but when $c=d$ (while $a\ne b$) we retain finite contributions. Moreover, under permutation of indices, the correlation function satisfies $\mathcal{H}^{ab}_{cd}(t) = -\mathcal{H}^{ba}_{cd}(t) = \mathcal{H}^{ab}_{dc}(t)  = -\mathcal{H}^{ba}_{dc}(t)$. We further recognize that $\mathcal{H}^{ab}_{cd}(0)=0$ and $\mathcal{H}^{ab}_{cd}(-t)=-\mathcal{H}^{ab}_{cd}(t)$. The expression in Eq. (\ref{eq:TWAPerturbativeResultHarmonic1}) serves as the starting point for the derivation of all subsequent perturbative expressions in this manuscript, and step-by-step instructions on how to proceed are provided in Appendix \ref{app:derivations}.

To proceed analytically, one must specify details about the initial density operator, the functional form of the perturbation operator defined in Eq. (\ref{eq:Perturbation}), and the operator whose time-dependent expectation value one is interested in computing. For the analyses presented here, although one can straightforwardly treat correlated initial conditions, we consider a factorized initial density, $\hat{\rho}(0) = \hat{\rho}_{\text{sys}}(0) \otimes \hat{\rho}_{\text{env}}(0)$, such that its partial Wigner transform assumes a Gaussian form and is given by
\begin{equation}\label{eq:BathWignerFunction}
W_{n_0,n'_0}(\mathbf{x}_{0},\mathbf{p}_{0}) = {\rho}^{(n_0,n'_0)}_{\text{sys}}(0) \ \prod_k \frac{\hbar}{\sigma_{x_k}\sigma_{p_k}}e^{-\frac{(x_{0_k}-x'_k)^2}{2\sigma_{x_k}^2} - \frac{(p_{0_k}-p'_k)^2}{2\sigma_{p_k}^2}},
\end{equation}
where $ {\rho}^{(n_0,n'_0)}_{\text{sys}}(0)=\bra{n_0}\hat{\rho}_{\text{sys}}(0)\ket{n'_0}$. We further impose that $\sigma_{p_k} = \omega_k \sigma_{x_k}$. This seemingly-restrictive specification encompasses two widely-used choices for initial densities where in thermal equilibrium \footnote{In thermal equilibrium we assume the form $\hat{\rho}_{\text{env}} = \frac{1}{Z_{\text{env}}}e^{-\beta\hat{H}_{\text{env}}}$ with $Z_{\text{env}}$ as the partition function, $\hat{H}_{\text{env}}$ as the bare environmental Hamiltonian, and $\beta = \frac{1}{k_BT}$.} 
one would impose that $\sigma_{x_k}=\sqrt{\frac{\hbar}{2\omega_k\tanh(\beta\hbar\omega_k/2)}}$ for all $k$, or to describe a minimum uncertainty wave packet centered at $(x'_k,p'_k)$ one would choose $\sigma_{x_k}=\sqrt{\frac{\hbar}{2\omega_k}}$. We note that polynomial terms (for example, the Laguerre polynomials used to describe particular harmonic eigenstates) in the initial distribution can be incorporated rather straightforwardly.\cite{AnalyticVibronicSpectra}

\section{Application to qubit decoherence and entanglement}
\label{sec:QubitDissipation}
Since the perturbative framework introduced here is amenable to expectation values in the \textit{combined} system and environment Hilbert space, it serves as a consistent, systematically improvable means of monitoring how decoherence is accompanied by system-environment entanglement formation. To this end, we consider a spin-boson model as a representation of a qubit embedded in a harmonic environment. The Hamiltonian is given by
\begin{equation}\label{eq:QubitHamiltonian}
\hat{H} = \frac{\epsilon}{2}\hat{\sigma}_z + \Delta \hat{\sigma}_x + \frac{1}{2} \sum_k \left(\hat{p}_k^2 + \omega_k^2 (\hat{x}_k - x_{0_k}\hat{\sigma}_z )^2\right),
\end{equation}
where $\{\hat{\sigma}_j\}$ are the usual Pauli spin operators, $\epsilon =  \epsilon_\uparrow - \epsilon_\downarrow=50 \ \text{cm}^{-1}$ is the detuning energy between the spin-up ($\uparrow$) and spin-down ($\downarrow$) states, and $\Delta = 10 \ \text{cm}^{-1}$ is the tunneling parameter. The qubit-environment coupling is governed by a spectral density of the Drude-Lorentz form,
\begin{equation}\label{eq:DrudeLorentz}
J(\omega) = 2\lambda\frac{\omega/\omega_c}{1+(\omega/\omega_c)^2},
\end{equation}
where $\lambda = 50 \ \text{cm}^{-1}$ is the total reorganization energy and $\omega_c = 100 \ \text{cm}^{-1}$ is the so-called cutoff frequency that is inversely proportional to the environmental correlation time. Here, the spectral density does not have system state-dependence, as it is taken to be coupled to the spin states in an anti-correlated fashion, proportional to $\hat{\sigma}_z$.

Within the present model the perturbation operator is environment independent and given by
\begin{equation}\label{eq:QubitPerturbation}
\hat{V} = \Delta\hat{\sigma}_x.
\end{equation}
As a result, we can factor $\Theta^{\{n_j,n'_j\}}_N\left(\left\{\big(\mathbf{\tilde{x}}_{\text{Bopp}}^{(j)}(\tau_j),\tau_j\big) \right\}\right) $ out of the integral over environmental phase space initial conditions and immediately evaluate the expression in Eq. (\ref{eq:TWAPerturbativeResultHarmonic1}) at $\delta p^{(k)}_{\tau_j} = 0$ for all environmental modes $k$ and perturbation operation times $\tau_j$. We then find that the $N$th order contribution to an expectation value is given by
\begin{equation}\label{eq:EVNoBath}
\begin{split}
\langle \hat{O}(t) \rangle_N = & \left( \frac{-i\Delta}{\hbar}\right)^N \left\{ \prod_{j=1}^N \int_0^{\tau_{j+1}}d\tau_j\right\}  \sum_{\{n_j,n'_j\}} \ e^{-\frac{i}{\hbar}\sum_{j=0}^N ({\epsilon}_{n_j}-{\epsilon}_{n'_j}) (\tau_{j+1} - \tau_j)} e^{  \Phi^{(N,\Hh)}_{\{n_j,n'_j\}}(\{\tau_j\}_{j=0}^{N+1})} \\
& \times  \left\{\prod_{j=1}^{N} \left(\sigma_x^{(n_{j}n_{j-1})}\delta_{n'_{j-1}n'_{j}} - \sigma_x^{(n'_{j-1}n'_{j})} \delta_{n_{j}n_{j-1}}\right) \right\} \\
& \times \int \frac{d\mathbf{x}_{0}d\mathbf{p}_{0}}{(2\pi\hbar)^D} \ O_W^{n'_N,n_N}(\mathbf{x}^{(N)}_{cl}(t),\mathbf{p}^{(N)}_{cl}(t))  W_{n_0,n'_0}(\mathbf{x}_{0},\mathbf{p}_{0}) \  \\
\times \ & e^{\frac{i}{\hbar} \sum_{k}  \left( \sum_{j=0}^N\deltaxk{n_j}{n'_j}[\sin(\omega_k \tau_{j+1}) - \sin(\omega_k \tau_j)]\right)\omega_k x_{0_k} }e^{- \frac{i}{\hbar} \sum_{k}  \left( \sum_{j=0}^N\deltaxk{n_j}{n'_j}[\cos(\omega_k \tau_{j+1}) - \cos(\omega_k \tau_j)]\right) p_{0_k} },
\end{split}
\end{equation}
where $\sigma_x^{(n_{j}n_{j-1})}=\bra{n_j}\hat{\sigma}_x \ket{n_{j-1}}$ is a Pauli spin operator matrix element.

In our analyses, we consider a zero-temperature initial density for the environment, as described in Eq. (\ref{eq:BathWignerFunction}), while assuming the spin system has initially been prepared in a symmetric linear combination of the spin-up and spin-down states. The initial density is then given by $\hat{\rho}(0) = \ket{\psi_0}\bra{\psi_0} \otimes \prod_k \ket{0_k}\bra{0_k}$ with $\ket{\psi_0} = \sqrt{\frac{1}{2}}(\ket{\uparrow} + \ket{\downarrow})$ so that the initial environmental Wigner function is given by Eq. (\ref{eq:BathWignerFunction}) with $(\mathbf{x'},\mathbf{p'})=(0,0)$ and $\sigma_{x_k} =\sqrt{\frac{\hbar}{2\omega_k}}$. Given our choice of pure state initial conditions for the environment, qubit decoherence is entirely driven by qubit-environment entanglement formation, which can be quantified in terms of a decay of the purity of the qubit reduced density matrix as $\text{Pur}(\rho_{\text{sys}}(t)) \equiv \text{Tr}[\rho_{\text{sys}}^2(t)] =\frac{1}{2}(1 + |a(t)|^2)$. Here, $\rho_{\text{sys}}(t)$ is the qubit (system) reduced density matrix and $|a(t)|^2 \equiv \langle \hat{\sigma}_x(t) \rangle^2 + \langle \hat{\sigma}_y(t) \rangle^2 + \langle \hat{\sigma}_z(t) \rangle^2$ is the squared length of the Bloch vector characterizing the state of the qubit. The purity reaches a value of $\frac{1}{2}$ for a maximally entangled state.

\begin{figure}[htb]
\includegraphics[width=\linewidth]{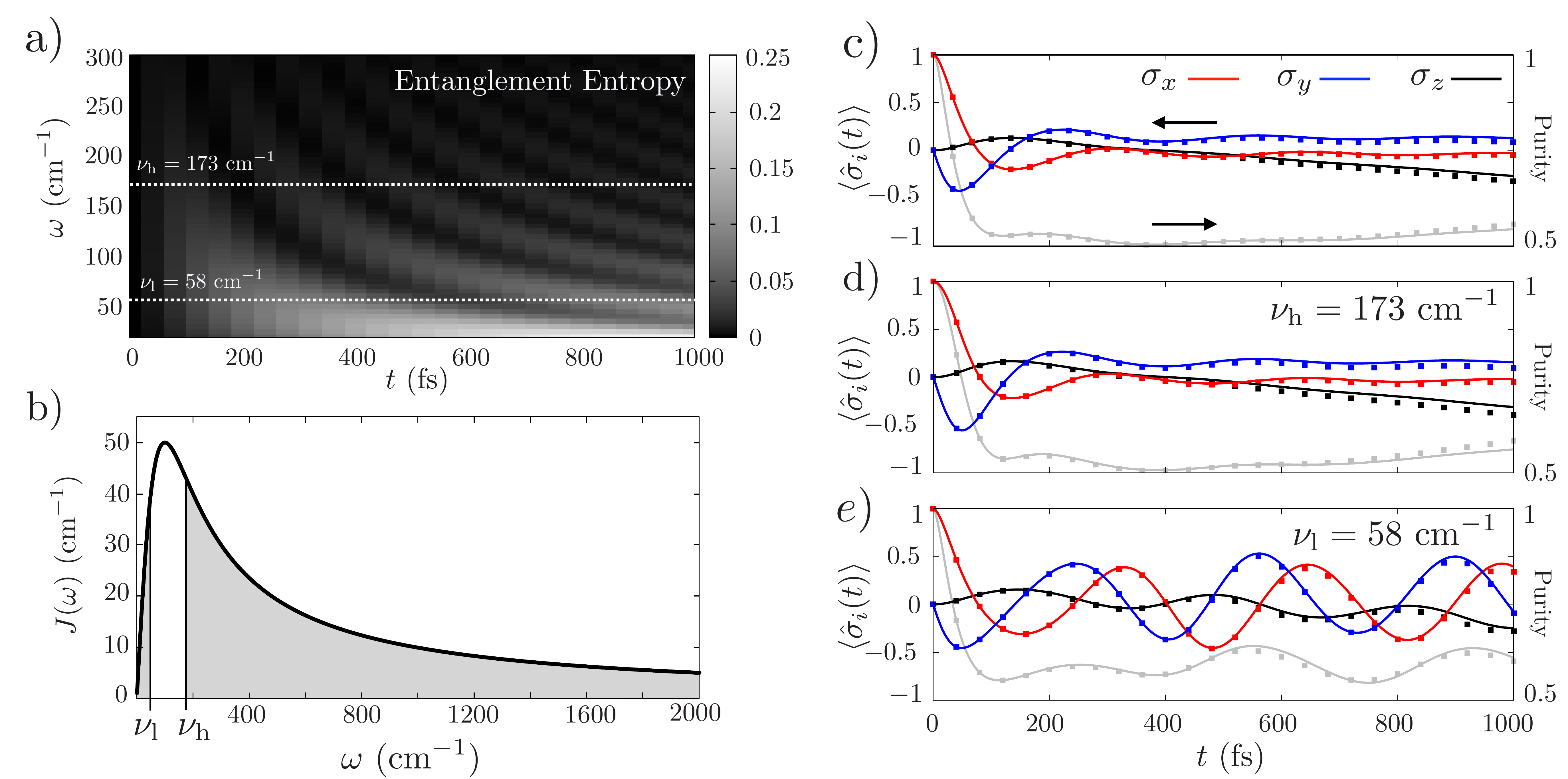}
\caption{Time evolution of a qubit coupled to a harmonic environment as described by our perturbative framework. In panel (a) we compute an environmental mode-specific time-dependent entanglement entropy as $S_k = - k_B \text{Tr}[\rho_{\text{env}}^{(k)}(t) \log \rho_{\text{env}}^{(k)}(t)]$ with $k_B = 1$. In panel (b) we illustrate the form of the spectral density describing qubit-environment coupling and highlight (through shaded regions) the spectral ranges where qubit-environment coupling may be suppressed in our applications. In panel (c) we characterize the time evolution of the qubit through spin operator expectation values, $\langle \hat\sigma_z(t)\rangle$ (black line), $\langle \hat\sigma_x(t)\rangle$ (red line), and $\langle \hat\sigma_y(t)\rangle$ (blue line). We also compute the purity of the qubit reduced density matrix as a measure of the entanglement that it forms with its environment (grey line). In panels (d) and (e) we present the time evolution of the qubit under a modified spectral density where certain environmental modes are completely decoupled to the qubit. In panel (d), all modes above $\nu_\text{h} = 173 \ \text{cm}^{-1}$ are decoupled, while in panel (e), all modes below $\nu_\text{l} = 58 \ \text{cm}^{-1}$ are decoupled. In panels (c-e), lines represent the second order approximation for the qubit dynamics, while the points represent the third order approximation.}
\label{fig:Qubit}
\end{figure}

The time evolution of the qubit can be fully characterized in terms of expectation values of the Pauli spin operators, which serve as a mapping of the time-evolving quantum state of the qubit onto the Bloch sphere. In Fig. \ref{fig:Qubit}, we present the time-evolution of the qubit and its environment, where the originally-factorized state is allowed to relax toward an equilibrium state with finite entanglement between the qubit and its environment. Fig. \ref{fig:Qubit}(a) shows the participation of individual environmental modes in the total entangled state, which is obtained by computing a single-mode entanglement entropy as $S_k = - k_B \text{Tr}[\rho_{\text{env}}^{(k)}(t) \log \rho_{\text{env}}^{(k)}(t)]$, where $\rho_{\text{env}}^{(k)}(t)$ is the reduced density matrix for the $k$th environmental mode. For our initial density, where each environmental mode is in its ground state, this quantity reports on deviation away from that initial state through the formation of entanglement between the $k$th environmental mode and all other DOFs in the coupled system. That is to say that, for each computation, we define a new partitioning for a bipartite system in which the $k$th environmental mode constitutes one subsystem, while the qubit and all other environmental modes constitute the other. However, the formation of entanglement between two independent environmental DOFs can only occur as mediated through mutual interactions with the qubit. 

In order to calculate the reduced density matrix of an environmental mode, $\rho_{\text{env}}^{(k)}(t)$, we compute expectation values of projection operators within a basis of harmonic oscillator eigenstates (including populations and coherences). For the simulations presented here, truncating the environmental basis at the second excited harmonic oscillator eigenstate turns out to be sufficient to characterize the time-evolved state of each mode under consideration. If one were to consider, for example, finite temperatures, stronger qubit-environment coupling, or lower frequency environmental modes, then the truncated Hilbert space would need to be expanded further. The reduced density matrix of an environmental mode can be computed through direct application of Eq. (\ref{eq:EVNoBath}) by using the Weyl symbol of the harmonic eigenstate projection operator for a mode with frequency $\omega$, as given by
\begin{equation}\label{eq:VibrationalWeyl}
\begin{split}
\left(\ket{n}\bra{m}\right)_W(x,p)&= \frac{2^{m+1}}{\sqrt{n!m!}} \sum_{j=0}^{n}\sum_{k=0}^{\min\{j,m\}} k! {n\choose j} {j\choose k} {m \choose k}\left(\frac{-1}{2}\right)^k  (a^*)^{n-k} a^{m-k}e^{-2a^*a}.
\end{split}
\end{equation}
Here, $a \equiv \sqrt{\frac{\omega}{2\hbar}} (x + i\frac{p}{\omega})$ and ${n \choose k} = \frac{n!}{k!(n-k)!}$. The detailed derivation of this expression is presented in the Supporting Information. Computing these reduced density matrix elements amounts to inserting this Weyl symbol, evaluated at the classical trajectory endpoints, into Eq. (\ref{eq:EVNoBath}) and analytically performing the simple Gaussian integrals over phase space initial conditions. Here, we have computed the environmental reduced density matrices at second order in perturbation theory. Note that each order in the perturbative expansion describes an additional transition between states of the qubit and is not related to the strength of the qubit's interactions with its environment. Because of this, our analysis provides an exact description of the environment's time-evolution in the absence of coupling between qubit states, regardless of the strength of coupling between the qubit and its environment.

The entanglement between the qubit and its \textit{entire} environment\cite{salamon2017entanglement} is understood in terms of the aforementioned purity of the qubit reduced density matrix. In Fig. \ref{fig:Qubit}(c), we present the time-dependent purity and spin operator expectation values (that fully characterize the qubit reduced density matrix). As is clear by the rapid decrease of purity and damping of the oscillatory behavior of spin operator expectation values, the qubit decoheres and becomes nearly maximally entangled with its environment on the $t \approx 100 \ \text{fs}$ timescale. The qubit expectation values have been computed up to third order in perturbation theory. Since the third order result is nearly identical to the second order result on this timescale, we safely assume that the second order expansion used for the environmental mode reduced density matrices is at least qualitatively accurate.

The mode-specific entanglement entropy presented in Fig. \ref{fig:Qubit}(a) reveals that the formation of entanglement between the qubit and its environment is dominated by low frequency modes. The formation of entanglement between the qubit and its environment leads to a rapid loss of the qubit's purity, which quickly approaches a maximally entangled state on the $\sim 100 \ \text{fs}$ timescale, as shown in Fig. \ref{fig:Qubit}(c). We note, however, that the qubit begins to recover purity as it relaxes toward its equilibrium state, which is simply a result of the equilibrium state's large projection onto the qubit spin-down state.

 As a simple demonstration of the potential impact of qubit-environment engineering, and to understand how different environmental DOFs contribute to the loss of qubit purity, we consider the cases where the qubit-environment coupling is selectively suppressed below and above certain threshold frequencies. We define a suppressed reorganization energy, $\tilde\lambda$, which we choose to be preserved among the two different environment engineering schemes. By completely decoupling the qubit from high frequency environmental DOFs, we express the suppressed reorganization energy as $\tilde{\lambda} = \frac{1}{\pi}\int_0^{\nu_{\text{h}}} d\omega \frac{J(\omega)}{\omega} = \frac{2\lambda}{\pi}\arctan(\nu_{\text{h}}/\omega_c)$
where the spectral density $J(\omega)$ is given in Eq. (\ref{eq:DrudeLorentz}). As a result, the qubit experiences an effective spectral density (defined for positive frequencies) of the form $\tilde{J}(\omega) = J(\omega)\Theta(\nu_{\text{h}}-\omega)$ with $\Theta(x)$ as the Heaviside step function. One can then extract the upper bound, $\nu_{\text{h}}$, for a given suppressed reorganization energy as 
\begin{equation}\label{eq:HighCut}
\nu_{\text{h}}=\omega_c\tan(\frac{\pi}{2}\alpha), 
\end{equation}
where $\alpha = \tilde{\lambda}/\lambda$. Similarly, we can define an effective spectral density where the low frequency environmental DOFs are decoupled as $\tilde{\lambda} = \frac{1}{\pi}\int_{\nu_{\text{l}}}^{\infty} d\omega \frac{J(\omega)}{\omega} = \lambda -  \frac{2\lambda}{\pi}\arctan(\nu_{\text{l}}/\omega_c)$ so that the effective spectral density can be expressed as $\tilde{J}(\omega) = J(\omega)\Theta(\omega-\nu_{\text{l}})$ and the lower bound, $\nu_{\text{l}}$, for a given suppressed reorganization energy is found as 
\begin{equation}\label{eq:LowCut}
\nu_{\text{l}}=\omega_c\tan(\frac{\pi}{2}(1-\alpha)).
\end{equation}

In our application, we have chosen $\alpha = \frac{2}{3}$ so that the suppressed reorganization energy is $\tilde{\lambda} \approx 33.33 \ \text{cm}^{-1}$. The resulting spectral density is depicted in Fig. \ref{fig:Qubit}(b) alongside the original version. In Figs. \ref{fig:Qubit}(d) and \ref{fig:Qubit}(e) we compute the time-evolution of the qubit in the presence of these engineered spectral densities. Since the reorganization energy is preserved between the two suppression schemes, differences in the qubit dynamics are simply a matter of which environmental modes are decoupled from the qubit rather than a relative difference in the total reorganization energy. In Fig. \ref{fig:Qubit}(d), where high frequency environmental DOFs above $\nu_{\text{h}}= 173 \ \text{cm}^{-1}$ are suppressed, we see that the purity is largely unchanged relative to the unsuppressed case. Notably, even though the reorganization energy is reduced by a third relative to the unsuppressed result, the only significant difference between the two cases occurs on the sub $\sim 50 \ \text{fs}$ timescale, where the high frequency (fast) modes are responsible for the early-time entanglement formation and the rate of purity loss in the suppressed simulation is slower than in the full simulation. In Fig. \ref{fig:Qubit}(e), however, where low frequency environmental DOFs below $\nu_\text{l}= 58 \ \text{cm}^{-1}$ are decoupled from the qubit, the overall decay in purity is significantly mitigated. This is presumably because, even though the range of suppressed environmental modes is smaller ($\omega \in (0,\nu_\text{l}]$ for the low frequency suppression compared to $\omega \in [\nu_{\text{h}},\infty)$ for the high frequency suppression), the lower frequency environmental modes are the dominant source of qubit-environment entanglement at timescales beyond $\sim 100 \ \text{fs}$. This much can be seen in Fig. \ref{fig:Qubit}(a), where the modes below $\nu_\text{l}$ have the largest entanglement entropy after $t \approx 100 \ \text{fs}$. Our analysis thus reveals that, while the high frequency environmental DOFs are responsible for the early-time formation of qubit-environment entanglement, they don't contribute substantially on the longer timescale, where lower frequency environmental DOFs are the dominant source of qubit decoherence. Concerted with the purity behavior, the decoherence is also significantly suppressed upon decoupling the low frequency environmental modes, as the qubit dynamics remains coherent throughout the entire timescale after an initial drop at short times. The comparative difference relative to the high frequency suppressed case indicates that the total reorganization energy does not represent a reliable metric for predicting qubit decoherence timescales. Rather, to understand the impact of qubit-environment coupling on different timescales, one must consider how the total reorganization energy is distributed among environmental modes of different frequencies.

\section{Environmental operator-dependent perturbations}
\label{sec:EnvironmentalDependence}

In the previous section, the perturbation operator was taken to be independent of environmental position operators and, because of this, the momentum fluctuation analysis was never utilized. Oftentimes, the coupling between system states cannot be chosen to be independent of environmental position operators. Alternatively, even if one is able to construct a model Hamiltonian that has environment-independent couplings, reformulating the problem in a basis that yields environment-dependent coupling may foster a perturbative expansion that converges more rapidly. In these cases, we can implement the momentum fluctuation analysis, as described in the Supporting Information, to fully treat the environmental operator dependence. In what follows, so that we can directly compare the environmental operator-dependent and environmental operator-independent perturbative analyses, we consider a simple basis rotation that allows us to perform these analyses in a fully consistent manner.

We first consider the Hamiltonian in a representation amenable to the environmental operator-independent perturbative expansion as
\begin{equation}\label{eq:HamiltonianHarmonicModel}
\begin{split}
\hat{H}& = \sum_{k} \frac{\hat{p}_k^2}{2} + \sum_n \left(\tilde\epsilon_n + \sum_{k} \frac{1}{2}\omega_k^2 \hat{x}_k^2 - \omega_k^2\xnotk{n}  \hat{x}_k  \right)\ket{n}\bra{n} + \sum_{m\ne n} \Delta_{nm}\ket{n}\bra{m}, \\
\end{split}
\end{equation}
so that one may quickly recognize the system state-dependent potential to be of the form given in Eq. (\ref{eq:HarmonicModel}). Naturally, the perturbation operator can be chosen as $\hat{V} = \sum_{n \ne m}^M \Delta_{nm}\ket{n}\bra{m}$, which contains no environmental dependence. We refer to this representation as the \textit{local} basis moving forward. 

Another choice of perturbation operator follows directly from a change of system basis. Within the eigenbasis of an isolated system Hamiltonian, $\hat{h} = \sum_{n}\tilde\epsilon_n\ket{n}\bra{n} + \sum_{n\ne m}\Delta_{nm}\ket{n}\bra{m}$, we find that
\begin{equation}\label{eq:HamiltonianHarmonicModelRedfield}
\hat{H} = \sum_{k}  \frac{\hat{p}_k^2}{2} + \sum_{\alpha} \left[\tilde{E}_\alpha +  \frac{1}{2}\omega_k^2 \hat{x}_k^2 - \omega_k^2 \hat{x}_k \xnotk{\alpha,\alpha}\right] \ket{\alpha}\bra{\alpha} - \sum_{\alpha \ne \beta}\left[\sum_k \omega_k^2\xnotk{\alpha,\beta} \hat{x}_k \right]\ket{\alpha} \bra{\beta},
\end{equation}
where Greek letters, \textit{i.e.}, $\alpha$ and $\beta$, label system eigenstates (where we note that the environmental reorganization energy is included in the diagonalized Hamiltonian), and $\xnotk{\alpha,\beta} = \sum_{n} c_n^\alpha c_n^{\beta *} \xnotk{n}$, where $c_n^\alpha$ is a wave function expansion coefficient from the eigenvectors of the isolated system Hamiltonian. Moving forward, we refer to this basis as the \textit{eigenbasis}. With this choice, we still obtain a system state-dependent potential of the form in Eq. (\ref{eq:HarmonicModel}), but now the natural choice for the perturbation operator is $\hat{V} =  - \sum_{\alpha \ne \beta}\left[\sum_k \omega_k^2\xnotk{\alpha,\beta} \hat{x}_k\right]\ket{\alpha} \bra{\beta}$, which is a linear function of the environmental position operators. We note that we are not restricted to linear environmental operator dependence and can, for example, apply arbitrary order polynomials in the bath operator in a consistent way by inserting Eq. (\ref{eq:ThetaGeneral}) into Eq. (\ref{eq:TWAPerturbativeResultHarmonic1}). 

As a concrete example, let us first consider the case where the operator whose expectation value we seek is simply the system projection operator, $\hat{O} = \ket{n'_N}\bra{n_N}$, so that, following integration over phase space initial conditions, we recover the $N$th order perturbative contribution to a reduced density matrix element in the local basis as
\begin{equation}\label{eq:FRETdensity2}
\begin{split}
\bra{n_N} & \hat{\rho}_{\text{sys}}^{(N)}(t)\ket{n'_N} = \left( \frac{-i}{\hbar}\right)^N \left\{ \prod_{j=1}^N \int_0^{\tau_{j+1}}d\tau_j\right\}  \sum_{\{n_j,n'_j\}_{j=0}^{N-1}} \rho^{(n_0,n'_0)}_{\text{sys}}(0) e^{-\frac{i}{\hbar}\sum_{j=0}^N ({\epsilon}_{n_j}-{\epsilon}_{n'_j}) (\tau_{j+1} - \tau_j)}  \\
&\times e^{ \Phi^{(N,\text{Tot})}_{\{n_j,n'_j\}}(\{\tau_j\}_{j=0}^{N+1})} \ \left\{\prod_{j=1}^{N} (\Delta_{n_{j}n_{j-1}}\delta_{n'_{j-1}n'_{j}} - \delta_{n_{j}n_{j-1}}\Delta_{n'_{j-1}n'_{j}}) \right\}  \\
&\times e^{\frac{i}{\hbar} \sum_{k}  \omega_k \left( \sum_{j=0}^{N}\deltaxk{n_j}{n'_j}[\sin(\omega_k \tau_{j+1}) - \sin(\omega_k \tau_j)]\right) x'_{k} }e^{- \frac{i}{\hbar} \sum_{k}  \left( \sum_{j=0}^{N}\deltaxk{n_j}{n'_j}[\cos(\omega_k \tau_{j+1}) - \cos(\omega_k \tau_j)]\right) p'_{k} }.
\end{split}
\end{equation}
This choice of operator is appropriate whenever one is interested in computing expectation values of operators that act solely on the Hilbert space of the system, since $\langle \hat{O}_{\text{sys}} (t)\rangle = \text{Tr}\{\hat{O}_{\text{sys}} \hat{\rho}(t)\} = \text{Tr}_{\text{sys}}\{\hat{O}_{\text{sys}}\hat{\rho}_{\text{sys}}(t)\}$, where $\hat{\rho}_{\text{sys}}(t) = \Tr_{\text{env}}\{\hat{\rho}(t)\}$ is the system reduced density operator.

In Eq. (\ref{eq:FRETdensity2}), we have used Eq. (\ref{eq:BathWignerFunction}) for the environmental Wigner function and we have defined
\begin{equation}\label{eq:PhiTot}
\begin{split}
 \Phi^{(N,\text{Tot})}_{\{n_j,n'_j\}}&(\{\tau_j\}_{j=0}^{N+1}) =  \Phi^{(N,\Hh)}_{\{n_j,n'_j\}}(\{\tau_j\}_{j=0}^{N+1}) + \Phi^{(N,\Gg)}_{\{n_j,n'_j\}}(\{\tau_j\}_{j=0}^{N+1}),
 \end{split}
 \end{equation}
 which describes the total influence of the environment on the time-evolved reduced density matrix element with
 \begin{equation}\label{eq:AllGCorrelationFunctions}
 \begin{split}
& \Phi^{(N,\Gg)}_{\{n_j,n'_j\}}(\{\tau_j\}_{j=0}^{N+1})  =  \frac{1}{\hbar} \sum_{j=0}^N \left[\Gg_{n_jn'_j}^{n_jn'_j}(\tau_{j+1} - \tau_j) -\Gg_{n_jn'_j}^{n_jn'_j}(0)\right] \\ 
& -\frac{1}{\hbar} \sum_{j=1}^{N}\sum_{l=0}^{j-1}\left\{\Gg_{n_ln'_l}^{n_jn'_j}(\tau_{j+1} - \tau_{l+1}) - \Gg_{n_ln'_l}^{n_jn'_j}(\tau_{j+1} - \tau_l) +\Gg_{n_ln'_l}^{n_jn'_j}(\tau_{j} - \tau_l) - \Gg_{n_ln'_l}^{n_jn'_j}(\tau_{j} - \tau_{l+1})\right\}.
\end{split}
\end{equation}
Here, we have defined additional correlation functions, $\mathcal{G}^{ab}_{cd}(t)$, that are also related to spectral densities as
\begin{equation}\label{eq:GCorrelationFunction}
\mathcal{G}^{ab}_{cd}(t)= \frac{2}{\pi} \int_0^\infty d\omega \ \frac{\sigma_{x}(\omega)\sigma_{p}(\omega)}{\hbar\omega^2}\left(J_{ac}(\omega) - J_{ad}(\omega)-J_{bc}(\omega) +J_{bd}(\omega)\right) \cos(\omega t),
\end{equation}
where $\sigma_{x}(\omega)$ and $\sigma_{p}(\omega)$ denote the continuous-frequency standard deviations for environmental phase space initial conditions. In contrast to the correlation function $\mathcal{H}^{ab}_{cd}(t)$ defined in Eq. (\ref{eq:HCorrelationFunction}), $\mathcal{G}^{ab}_{cd}(t)$ is zero any time $a=b$ \textit{or} $c=d$ and we note that $\mathcal{G}^{ab}_{cd}(t) = -\mathcal{G}^{ba}_{cd}(t) = - \mathcal{G}^{ab}_{dc}(t) = \mathcal{G}^{ba}_{dc}(t)$ with $\mathcal{G}^{ba}_{cd}(-t)=\mathcal{G}^{ba}_{cd}(t)$  and, in general, $\mathcal{G}^{ab}_{cd}(0)\ne 0$. Moreover, these correlation functions depend on the details of the initial environmental density through the widths of the position and momentum distributions defined in Eq. (\ref{eq:BathWignerFunction}). In total, the $N$th order perturbative contribution to the time-evolved reduced density matrix element is computed by evaluating the contributing correlation functions at time delays between perturbative operations, and is proportional to the product of perturbation matrix elements as given in Eq. (\ref{eq:FRETdensity2}). Finally, the midpoints of the Gaussian Wigner functions describing the initial environmental density contribute according to the phase factor in Eq. (\ref{eq:FRETdensity2}).

While the reduced density matrix for this model in the local basis can be written down immediately to arbitrary order in the expansion (see Eq. (\ref{eq:FRETdensity2})), the eigenbasis requires an order-specific derivation (although one could likely generalize the result to arbitrary order). This is simply because the phase space initial condition integrand now contains a polynomial that is unique to each order in the expansion. Nonetheless, the derivation of reduced density matrix elements within the eigenbasis closely mimics that of the local basis. Proceeding analogously to the local basis result, we immediately arrive at Eq. (\ref{eq:TWAPerturbativeResultHarmonic1}) upon the substitution $(n_j,n'_j)\rightarrow(\alpha_j,\alpha'_j)$. In fact, Eq. (\ref{eq:TWAPerturbativeResultHarmonic1}) is general for arbitrary choices of the system basis. In the present case, it is given by
\begin{equation}\label{eq:TWAPerturbativeResultHarmonicRedfield}
\begin{split}
\langle \hat{O}(t)  \rangle_N & = \left( \frac{-i}{\hbar}\right)^N \left\{ \prod_{j=1}^N \int_0^{\tau_{j+1}}d\tau_j\right\}  \sum_{\{\alpha_j,\alpha'_j\}} \  e^{-\frac{i}{\hbar}\sum_{j=0}^N ({E}_{\alpha_j}-{E}_{\alpha'_j}) (\tau_{j+1} - \tau_j)} \\
& \times e^{ \Phi^{(N,\Hh)}_{\{\alpha_j,\alpha'_j\}}(\{\tau_i\}_{i=0}^{N+1}) } \int \frac{d\mathbf{x}_{0}d\mathbf{p}_{0}}{(2\pi\hbar)^D} \ W_{\alpha_0,\alpha'_0}(\mathbf{x}_{0},\mathbf{p}_{0}) \\
& \times \bigg[ \Theta^{\{\alpha_j,\alpha'_j\}}_N\left(\left\{\big(\mathbf{\tilde{x}}_{\text{Bopp}}^{(j)}(\tau_j)\big) \right\}\right)  \ O_W^{\alpha'_N,\alpha_N}(\mathbf{x}^{(N)}_{\delta p}(t),\mathbf{p}^{(N)}_{\delta p}(t)) e^{ - \frac{i}{\hbar} \phi_N(\mathbf{\delta p}) }\bigg]_{\delta p = 0} \\
& \times  e^{\frac{i}{\hbar} \sum_{k}  \left( \sum_{j=0}^N\deltaxk{\alpha_j}{\alpha'_j}[\sin(\omega_k \tau_{j+1}) - \sin(\omega_k \tau_j)]\right)\omega_k x_{0_k} }e^{- \frac{i}{\hbar} \sum_{k}  \left( \sum_{j=0}^N\deltaxk{\alpha_j}{\alpha'_j}[\cos(\omega_k \tau_{j+1}) - \cos(\omega_k \tau_j)]\right) p_{0_k} },
\end{split}
\end{equation}
where  ${E}_{\alpha} = \tilde{E}_\alpha - \frac{1}{2}\sum_{k}\omega_k^2 x_{0_k}^{(\alpha,\alpha) 2}$ and the correlation functions contributing to $ \Phi^{(N,\Hh)}_{\{\alpha_j,\alpha'_j\}}(\{\tau_j\}_{j=0}^{N+1})$, denoted as  $\mathcal{H}^{\alpha\beta}_{\gamma\delta}(t)$, are of the same form as Eq. (\ref{eq:HCorrelationFunction}), but where the spectral densities are now given by $J_{\alpha\beta}(\omega) = \frac{\pi}{2}\sum_k \omega_k^3 x_{0_k}^{(\alpha,\alpha)}x_{0_k}^{(\beta,\beta)}\delta(\omega-\omega_k)$ with $\xnotk{\alpha,\beta} = \sum_{n,m} c_n^\alpha c_m^{\beta \ *} \xnotk{n,m}$. We further define $\xbark{\alpha}{\beta} = \frac{1}{2} (\xnotk{\alpha,\alpha} + \xnotk{\beta,\beta})$ and $\deltaxk{\alpha}{\beta} = \xnotk{\alpha,\alpha} - \xnotk{\beta,\beta}$ as the eigenbasis mean and difference potential minima, respectively.

To make connection with the local basis reduced density matrix given in Eq. (\ref{eq:FRETdensity2}), we consider an expansion of the expectation value of the system projection operator in the eigenbasis, $\hat{O} = \ket{\alpha'_N}\bra{\alpha_N}$, corresponding to the time-evolved reduced density matrix element, $\bra{\alpha_N}\hat{\rho}_{\text{sys}}(t)\ket{\alpha'_N}$, and we again assume the environmental Wigner function is given by the form in Eq. (\ref{eq:BathWignerFunction}). With our choice of operator, the momentum fluctuation derivatives in Eq. (\ref{eq:TWAPerturbativeResultHarmonicRedfield}) only operate on the phase factor, $\phi_N(\mathbf{\delta p})$, and the shifted classical trajectories that appear as arguments in the $\Theta_N$ tensor. The result for the $N$th order contribution to the projection operator expectation value is obtained after taking the appropriate derivatives, evaluating the resulting expression at $\delta p^{(k)}_{\tau_j}=0$ for all modes $k$ and times $\tau_j$, and then solving the integrals over phase space initial conditions, which are simply polynomials multiplied by a Gaussian and which can therefore be solved analytically. Here, we derive explicit expressions up to second order in the perturbative expansion.

At $0$th order, where the perturbation does not contribute and we only get contributions when $(\alpha_0,\alpha'_0) = (\alpha_N,\alpha'_N)$, one simply finds
\begin{equation}\label{eq:RedfieldZeroth}
\begin{split}
& \bra{\alpha_N}\hat{\rho}^{(0)}_{\text{sys}}(t)\ket{\alpha'_N} \\
& = \rho^{(\alpha_N,\alpha'_N)}_{\text{sys}}(0)e^{-\frac{i}{\hbar}({E}_{\alpha_N} - {E}_{\alpha'_N})t} e^{\Phi^{(0,\text{Tot})}_{\alpha_N,\alpha'_N}(t)} e^{\frac{i}{\hbar} \sum_{k}  \omega_k \deltaxk{\alpha_N}{\alpha'_N}\sin(\omega_k t) x'_{k} }e^{- \frac{i}{\hbar} \sum_{k}  \deltaxk{\alpha_N}{\alpha'_N}[\cos(\omega_k t) - 1] p'_{k} },
\end{split}
\end{equation}
where we remind the reader that $(x'_k,p'_k)$ describes the center for a minimum uncertainty wave packet describing the initial state of the $k$th environmental DOF. 

The linear order ($N=1$) contribution, where the perturbation acts only at time $\tau_1$, can be found following simple Gaussian integration as 
\begin{equation}\label{eq:RedfieldFirst}
\begin{split}
 \bra{\alpha_N}\hat{\rho}^{(1)}_{\text{sys}}(t)\ket{\alpha'_N}=   \frac{-i}{\hbar} \int_0^{t}d\tau_1  \sum_{\alpha_0,\alpha'_0} \  \rho^{(\alpha_0,\alpha'_0)}_{\text{sys}}(0) & \ \chi^{(1)}_{\{\alpha_j\alpha'_j\}}(\{\tau_j\}_{j=0}^{2}) \   e^{-\frac{i}{\hbar} \sum_{j=0}^1({E}_{\alpha_{j}}-{E}_{\alpha'_{j}}) (\tau_{j+1} - \tau_{j})} \\
\times \ e^{\Phi^{(1,\text{Tot})}_{\{\alpha_j,\alpha'_j\}}(\{\tau_j\}_{j=0}^{2})} & e^{\frac{i}{\hbar} \sum_{k}  \omega_k \left( \sum_{j=0}^{1}\deltaxk{\alpha_j}{\alpha'_j}[\sin(\omega_k \tau_{j+1}) - \sin(\omega_k \tau_j)]\right) x'_{k} } \\
& e^{- \frac{i}{\hbar} \sum_{k}  \left( \sum_{j=0}^{1}\deltaxk{\alpha_j}{\alpha'_j}[\cos(\omega_k \tau_{j+1}) - \cos(\omega_k \tau_j)]\right) p'_{k} },
\end{split}
\end{equation}
where we note that $(\alpha_1,\alpha'_1) = (\alpha_N,\alpha'_N)$ and $\tau_2 = t$ in the expression above.  The (now explicitly time-dependent) Liouville pathway weight at linear order is found to be
\begin{equation}\label{eq:RedfieldLinearLiouville}
\begin{split}
  \chi^{(1)}_{\{\alpha_j\alpha'_j\}}&(\{\tau_j\}_{j=0}^{2}) = \\
  & -\sum_k \omega_k^2[\xnotk{\alpha_N,\alpha_0}\delta_{\alpha'_0,\alpha'_N} -\delta_{\alpha_N,\alpha_0} \xnotk{\alpha'_0,\alpha'_N}]   \left(\theta_{1,1}^{(k)}(\{\tau_j\}_{j=0}^{2}) - \frac{1}{2}S_1 \zeta_{1,1}^{(k)}(\{\tau_j\}_{j=1}^{2})\right).
\end{split}
\end{equation}
Here, the sign function that is determined by whether the perturbative operation acts on the left or right is given for the purely off-diagonal perturbation operators considered here as $S_j=(\delta_{\alpha'_{j-1}\alpha'_j} - \delta_{\alpha_{j}\alpha_{j-1}})$. The terms contributing to Eq. (\ref{eq:RedfieldLinearLiouville}) are given, in their general form, by
\begin{equation}\label{eq:LiouvilleTheta}
\begin{split}
\theta^{(k)}_{j,N}&(\{\tau_l\}_{l=0}^{N+1}) =  \bigg(\frac{i}{\hbar} \sigma_{x_k}\sigma_{p_k}\sum_{l=0}^N\deltaxk{\alpha_l}{\alpha'_l}[\sin(\omega_k(\tau_{l+1}-\tau_{j})) - \sin(\omega_k(\tau_{l}-\tau_{j}))] \\
&\hspace{3cm} + x'_k\cos(\omega_k\tau_{j}) + \frac{p'_k}{\omega_k}\sin(\omega_k\tau_{j}) + x^{(j-1)}_{k,nl}(\tau_{j})\bigg), 
\end{split}
\end{equation}
which is a result of the classical trajectory component of the $\Theta_N$ tensor's argument in Eq. (\ref{eq:TWAPerturbativeResultHarmonicRedfield}), with the term 
\begin{equation}\label{eq:Xnonlocal}
x_{k,nl}^{(N)}(t) =\sum_{j=0}^{N} \bar{x}_{0_k}^{\alpha_j \alpha'_j}[\cos(\omega_k (t-\tau_{j+1}))-\cos(\omega_k (t-\tau_j))].
\end{equation}
As a direct result of the derivative operation acting on the momentum fluctuation-dependent phase factor,  $\phi_N(\mathbf{\delta p})$, in Eq. (\ref{eq:TWAPerturbativeResultHarmonicRedfield}), we have that
\begin{equation}\label{eq:LiouvilleZetaj}
 \zeta_{j,N}^{(k)}(\{\tau_l\}_{l=j}^{N+1}) =  \sum_{l=j}^N\deltaxk{\alpha_l}{\alpha'_l}[\cos(\omega_k(\tau_{l+1}-\tau_j)) - \cos(\omega_k(\tau_l-\tau_j))].
 \end{equation}

At second order ($N=2$), one readily finds
\begin{equation}\label{eq:RedfieldSecond}
\begin{split}
 \bra{\alpha_N}\hat{\rho}^{(2)}_{\text{sys}}(t)\ket{\alpha'_N}=   \frac{-1}{\hbar^2} \int_0^{t}d\tau_2 \int_0^{\tau_2}d\tau_1 \sum_{\substack{\alpha_0,\alpha'_0 \\ \alpha_1,\alpha'_1 }} & \ \rho^{(\alpha_0,\alpha'_0)}_{\text{sys}}(0) \ \chi^{(2)}_{\{\alpha_j\alpha'_j\}}(\{\tau_j\}_{j=0}^3) \\
 \times \ e^{-\frac{i}{\hbar} \sum_{j=0}^2({E}_{\alpha_{j}}-{E}_{\alpha'_{j}}) (\tau_{j+1} - \tau_{j})} e^{\Phi^{(2,\text{Tot})}_{\{\alpha_j,\alpha'_j\}}(\{\tau_j\}_{j=1}^{3})} & e^{\frac{i}{\hbar} \sum_{k}  \omega_k \left( \sum_{j=0}^{2}\deltaxk{\alpha_j}{\alpha'_j}[\sin(\omega_k \tau_{j+1}) - \sin(\omega_k \tau_j)]\right) x'_{k} } \\
\times \ & e^{- \frac{i}{\hbar} \sum_{k}  \left( \sum_{j=0}^{2}\deltaxk{\alpha_j}{\alpha'_j}[\cos(\omega_k \tau_{j+1}) - \cos(\omega_k \tau_j)]\right) p'_{k} }
\end{split}
\end{equation}
with $(\alpha_2,\alpha'_2) = (\alpha_N,\alpha'_N)$, $\tau_3 = t$, and where the second order Liouville pathway weight is given by
\begin{equation}\label{eq:RedfieldSecondLiouville}
\begin{split}
  \chi^{(2)}_{\{\alpha_i\alpha'_i\}_{i=0}^2}&(t,\tau_2,\tau_1) =  \sum_{k,k'} \omega_k^2\omega_{k'}^2\bigg[x_{0_{k'}}^{(\alpha_N,\alpha_1)}x_{0_{k}}^{(\alpha_1,\alpha_0)}\delta_{\alpha'_0,\alpha'_1}\delta_{\alpha'_1,\alpha'_N} - x_{0_{k'}}^{(\alpha_N,\alpha_1)}\delta_{\alpha_1,\alpha_0}
 x_{0_{k}}^{(\alpha'_0,\alpha'_1)}\delta_{\alpha'_1,\alpha'_N} \\
 & \hspace{3.0cm} - \delta_{\alpha_N,\alpha_1}x_{0_{k}}^{(\alpha_1,\alpha_0)}\delta_{\alpha'_0,\alpha'_1}x_{0_{k'}}^{(\alpha'_1,\alpha'_N)} + \delta_{\alpha_N,\alpha_1} \delta_{\alpha_1,\alpha_0}x_{0_{k}}^{(\alpha'_0,\alpha'_1)}x_{0_{k'}}^{(\alpha'_1,\alpha'_N)}\bigg]\\
 &\times \bigg[ \left(\theta_{2,2}^{(k')}(\{\tau_j\}_{j=0}^3) - \frac{1}{2}S_2 \zeta_{2,2}^{(k')}(\{\tau_j\}_{j=2}^3)\right)\left(\theta_{1,2}^{(k)}(t\{\tau_j\}_{j=0}^3) - \frac{1}{2}S_1 \zeta_{1,2}^{(k)}(\{\tau_j\}_{j=1}^3)\right) \\
 & \hspace{3cm} + \left(\frac{\sigma_{x_k}\sigma_{p_k}}{\omega_k}\cos(\omega_k(\tau_2 - \tau_1)) - \frac{i\hbar}{2\omega_k}S_1\sin(\omega_k(\tau_2-\tau_1))\right) \delta_{k,k'}\bigg].
\end{split}
\end{equation}
This analysis can be systematically applied to higher order perturbations, but we restrict ourselves here to the second order expansion.

The Liouville pathway weights, $ \chi^{(1)}_{\{\alpha_j\alpha'_j\}}(\{\tau_j\}_{j=0}^2)$ and $\chi^{(2)}_{\{\alpha_j\alpha'_j\}}(\{\tau_j\}_{j=0}^3)$, can be represented entirely in terms of two-point autocorrelation functions related to generalized environmental spectral densities. This has the distinct computational advantage of allowing one to calculate these correlation functions one time, and then simply evaluate them at different time delays while performing the necessary temporal integrals in Eqs. (\ref{eq:RedfieldFirst}) and (\ref{eq:RedfieldSecond}) numerically. The generalized spectral densities defining the correlation functions are proportional to the delocalization of the system-bath coupling strength between eigenstates, and the Liouville pathway weights can be expressed in terms of them as demonstrated in Appendix \ref{app:correlations}.

We note that, even though these expressions are derived for $N$th order in perturbation theory, they generally contain terms up to $2N$th order in system-bath coupling due to the explicit equations-of-motion of the environmental modes as well as the result of Gaussian integration over phase space initial conditions. Moreover, we emphasize that the preceding analysis is not in any way restricted to the eigenbasis and rather can be considered as a general method for incorporating linear order environmental operator dependence in the coupling between system states in any basis. In fact, it would be quite straightforward to extend this analysis to account for perturbation matrix elements that have higher order dependence on environmental position operators using the momentum fluctuation analysis presented in the Supporting Information and subsequently performing the necessary Gaussian integrals.

\begin{figure}[htb]
\includegraphics[width=0.75\linewidth]{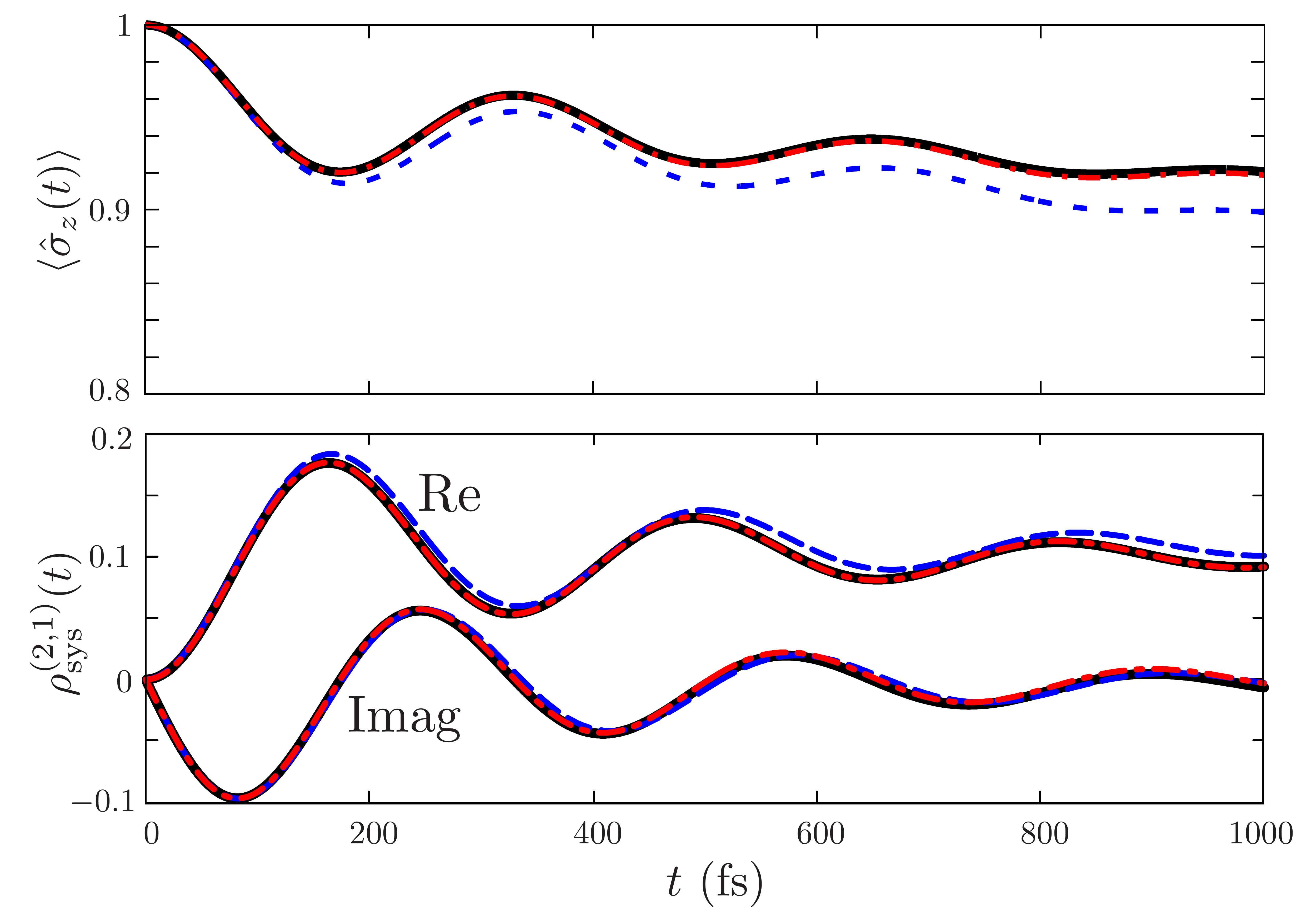}
\caption{Time-evolution of reduced density matrix elements for a model spin-boson system with weak system-environment coupling. In the top panel, we present the time-dependent expectation value of the operator $\hat{\sigma}_z = \ket{1}\bra{1} - \ket{2}\bra{2}$. In the bottom panel, we present the real and imaginary parts of the $\rho^{(2,1)}_{\text{sys}}(t)$ reduced density matrix element. In all cases, the solid black line corresponds to the numerically exact HEOM result, the dashed blue line corresponds to the second order local basis expansion result, and the dash-dotted red line corresponds to the second order eigenbasis expansion result.}
\label{fig:SpinBoson}
\end{figure}

In Fig. \ref{fig:SpinBoson} we demonstrate the application of each of the described perturbative expansions in a parameter regime where one would expect reasonable agreement with exact results, due to the existence of a well-defined smallness parameter. We consider a spin-boson model with an energy gap of $\epsilon_1 - \epsilon_2= 100 \ \text{cm}^{-1}$ and a coupling of $\Delta_{12}=\Delta_{21} = 10 \ \text{cm}^{-1}$. Each state is coupled weakly to its own independent harmonic environment described by a Drude-Lorentz spectral density, as given in Eq. (\ref{eq:DrudeLorentz}), with $\omega_c = 53.08 \ \text{cm}^{-1}$ and $\lambda = 1 \ \text{cm}^{-1}$. Such a model is appropriate for describing a donor-acceptor pair where each are weakly coupled to their environments. We assume that the initial density is of factorized form, and that the environmental Wigner function is thermal with $(\mathbf{x'},\mathbf{p'}) = (0,0)$ and $\sigma_{x_k}=\sqrt{\frac{\hbar}{2\omega_k\tanh(\beta\hbar\omega_k/2)}}$ for all $k$ at a temperature of $T = 300 \ \text{K}$.\cite{trushechkin2019calculation} Because of the weak environmental coupling, the eigenbasis expansion outperforms its local basis counterpart when compared to the numerically exact HEOM result.\cite{strumpfer2012open} However, the local basis result still provides a good description of the time-evolved density matrix even as $\lambda \ll \Delta_{12}$.\cite{trushechkin2019calculation} 

\begin{figure}[htb]
\includegraphics[width=0.75\linewidth]{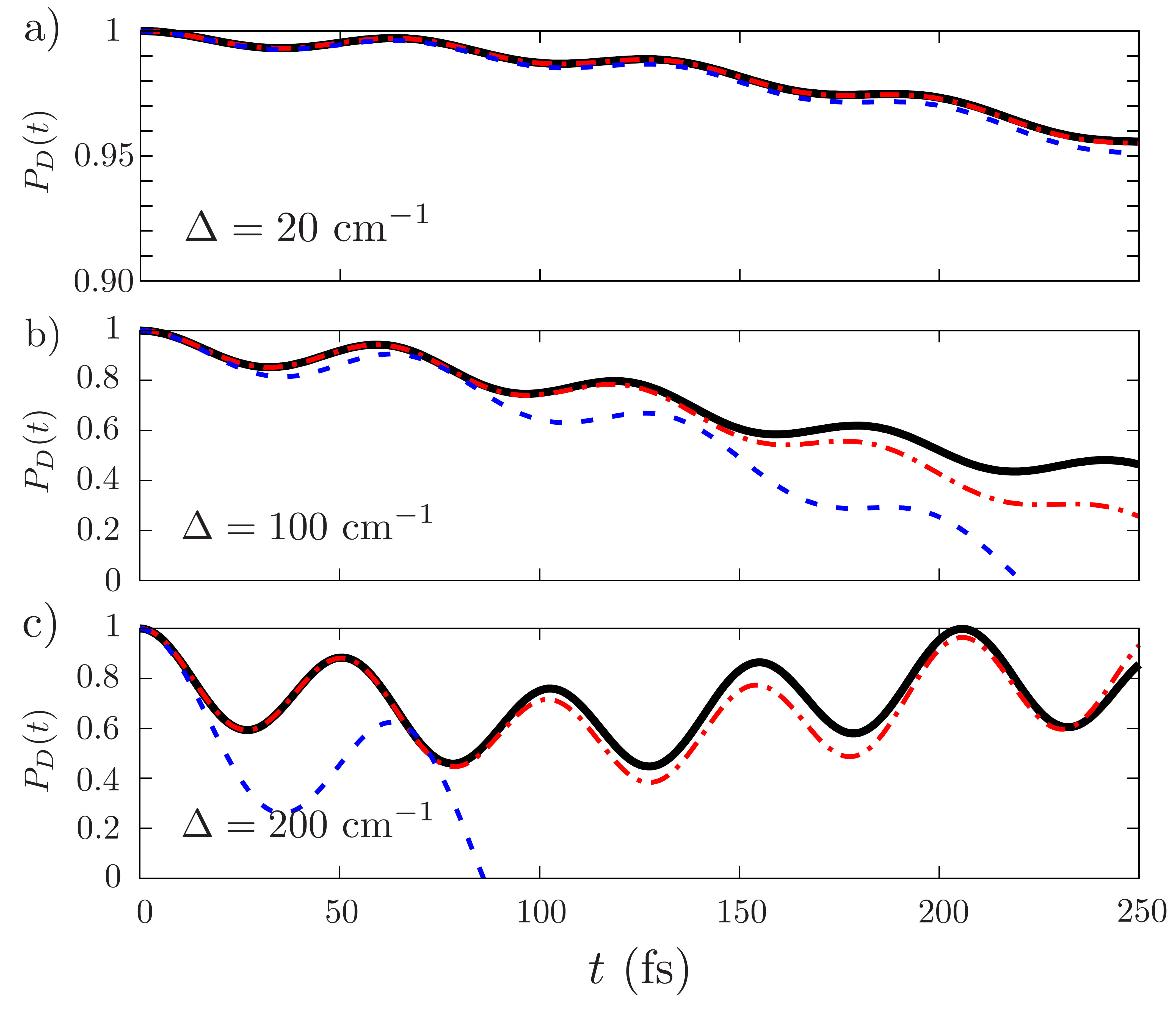}
\caption{Time-dependent donor population for a donor-acceptor pair with a single vibrational mode coupled to the acceptor state for various choices of the donor-acceptor coupling strength. All other parameters are described in the main text. In each plot, the numerically exact HEOM result is given by the solid black line, the dashed blue line corresponds to the second order local basis expansion result, and the dash-dotted red line corresponds to the second order eigenbasis expansion result.}
\label{fig:Resonant}
\end{figure}

As previously mentioned, the described framework does not inherently make any assumptions about the decay times of the environmental correlation functions that are central to the derived equations. As such, one is not restricted to unstructured environmental spectral densities; something we have already exploited when modifying the spectral density in Fig. \ref{fig:Qubit}(b). For demonstrative purposes, we proceed by considering the limiting case of a single (discrete) environmental mode that is coupled to the acceptor state of a two-level system at $T = 300 \ \text{K}$. In this application, we consider the mode to have a frequency of $\omega_0 = 500 \ \text{cm}^{-1}$, a reorganization energy of $\lambda = 25 \ \text{cm}^{-1}$, and the local basis energy gap is given as ${\epsilon}_1 - {\epsilon}_2 = \hbar \omega_0$ so as to be resonant with the single environmental mode. In Fig. \ref{fig:Resonant} we plot the time-dependent donor (state 1) population for multiple choices of the tunneling parameter, $\Delta$, scanning from a small value, where the local basis perturbative expansion provides good agreement with exact results, to a large value, where this expansion breaks down at second order due to secular divergences, while the eigenbasis perturbative expansion continues to provide qualitative agreement. 

\section{Conclusions and Outlook}
\label{sec:conclusions}
In this manuscript, we have introduced a framework for combining the truncated Wigner approximation with perturbative calculations of time-evolved expectation values of operators for a discrete quantum system in contact with an environment. For the case where the environment is harmonic and where the system state dependent harmonic potentials differ at most by a relative shift of the potential minima, we have demonstrated how one can utilize the described framework to derive expressions for the $N$th order perturbative contribution to operator expectation values in the combined system-environment Hilbert space, even including polynomial bath operator dependence in the perturbative operator. In contrast to dynamical methods based on projection operator techniques\cite{breuer2002theory,hwang2015coherent,geva2021electronic,mulvihill2021road,montoya2015extending} or influence functionals\cite{tanimura2020numerically,makri1995numerical,diosi1997non}, our framework retains full dynamical information for the system \textit{and} its environment while also avoiding common assumptions about the initial environmental density.

Our framework is of particular interest to quantum information applications, which require a detailed understanding of how quantum decoherence is driven by a growth of system-environment entanglement. \cite{rossatto2011purity,pernice2012system,roszak2018criteria,roszak2015characterization,costa2016system} We have highlighted the potential of our framework to unravel such processes by computing decoherence of a qubit at zero temperature while demonstrating a concerted entanglement growth of individual environmental modes. By analyzing the qubit-environment entanglement behavior, we identify the low-frequency modes to particularly drive decoherence and a specific suppression of their interactions considerably suppresses decoherence compared to the selective suppression of high-frequency modes at a constant environmental reorganization energy. This indicates that reorganization energy itself as a relevant metric for environment-induced decoherence is incomplete. We believe this finding to exemplify that our framework can reveal important guiding principles for optimizing qubit coherence times. Such an analysis can trivially be extended to realistic structured environmental spectral densities, obtained directly from experimental measurements, for which the strongest-entangled environmental modes can be readily identified. It should be noted that other factors have been suggested to drive quantum decoherence beyond system-bath entanglement,\cite{eisert2002quantum,hilt2009system,pernice2011decoherence} something that would also be of interest to explore with the framework presented here.

We have furthermore demonstrated the ability of our framework to treat reduced density matrix dynamics under a Hamiltonian where a simple basis rotation allows us to interpolate between an environment-dependent and environment-independent perturbation. In doing so, bath operator dependent perturbations are described through a momentum fluctuation analysis related to the so-called Bopp operators as derived within a phase space path integral formulation, which is a general result for perturbation operators that are polynomials in environmental operators.\cite{polkovnikov2010phase,bopp1961statistische,AnalyticVibronicSpectra} The flexibility of our method to include polynomial-type perturbation operators allows for the investigation of system-environment interactions beyond the commonly-adopted bi-linear form.

While we stress that the framework described here is not restricted to deriving reduced density matrix dynamics methods, we note that a direct application of the naive perturbation theory utilized here is well known to suffer from secular divergences that grow as $\sim t^{N}$ at long times.\cite{langhoff1972aspects} This does not negate the described framework, however, as one may apply common techniques to diminish their impact,\cite{reichman1996relaxation} or potentially derive iterative algorithms utilizing the described framework for short-time propagation segments -- something that is awarded by the perturbation theory's retention of the dynamical evolution of the environment. For harmonic models, one could imagine achieving this though a Monte Carlo evaluation of intermediate sums over system states, which can provide an efficient means to reconstruct the exact time-evolved reduced density matrix.\cite{huo2012consistent,provazza2018semiclassical} Alternatively, one could consider implementing the framework described here to compute short-time expectation values where one can efficiently perform the numerical integrals at a sufficiently high order in the perturbative expansion, and then utilize a non-Markovian transfer tensor method to extend the results to longer times. \cite{cerrillo2014non,rosenbach2016efficient,kananenka2016accurate,buser2017initial}

{\appendix
\section{Utilizing Eq. (\ref{eq:TWAPerturbativeResultHarmonic1}) to derive perturbative expressions}
\label{app:derivations}
In order to utilize Eq. (\ref{eq:TWAPerturbativeResultHarmonic1}) to derive perturbative contributions to the expectation value of an operator, $\hat{O}$, the procedure is as follows:
\begin{enumerate}
  \item Obtain the Weyl symbol of the operator $\hat{O}$. This can be achieved directly through the integral transformation as given by Eq. (\ref{eq:PartialWigner}) or by evaluating the normal ordered operator in terms of Bopp operators. \cite{polkovnikov2010phase} The Weyl symbol should then be evaluated along the \textit{shifted} environmental trajectories given by Eq. (\ref{eq:ShiftedTrajectory}).
  \item Evaluate the contributing tensor elements of $\Theta^{\{n_j,n'_j\}}_N\left(\left\{\big(\mathbf{\tilde{x}}_{\text{Bopp}}^{(j)}(\tau_j),\tau_j\big) \right\}\right) $ according to Eq. (\ref{eq:ThetaGeneral}), where the Bopp operators are defined in Eq. (\ref{eq:BoppOpp}). For perturbations that include environmental position operator dependence, the resulting expression will contain derivatives with respect to the infinitesimal fluctuations of the conjugate momentum, $\mathbf{\delta p}$. Any functional dependence on these momentum fluctuations will be provided by later-time perturbative interactions, the Weyl symbol of the operator-of-interest, and the phase factor $\phi_N(\mathbf{\delta p})$ defined in Eq. (\ref{eq:MomentumFluctuationPhase}).
  \item If the perturbation operator contains environmental position operator dependence, perform all momentum fluctuation derivatives and evaluate the resulting expression in the limit that all momentum fluctuations go to zero. One should now obtain an explicit function of environmental initial conditions, $f(\mathbf{x}_0, \mathbf{p}_0)$, and geometric parameters directly related to details of the shifted harmonic potentials that contribute along the series of system states connecting the endpoints of the propagation. The functional form of $f(\mathbf{x}_0, \mathbf{p}_0)$ depends on the form of the Weyl symbol of the operator-of-interest, the order of perturbation theory under consideration, and the functional form of the perturbation operator. Oftentimes, however, $f(\mathbf{x}_0, \mathbf{p}_0)$ can be simplified considerably through simple trigonometric identities.
 \item Perform the integrals over all environmental phase space initial conditions. These integrals will amount to a convolution of $f(\mathbf{x}_0, \mathbf{p}_0)$ from Step 3 with the Gaussian Wigner function in Eq. (\ref{eq:BathWignerFunction}) that describes the initial density of environmental modes along with the remaining phase factors given in Eq. (\ref{eq:TWAPerturbativeResultHarmonic1}).
\end{enumerate}

\section{Correlation functions for environment-dependent expansion}
\label{app:correlations}
We start by defining a set of environmental correlation functions as
\begin{align}
\mathcal{I}^{ab}_{cd}(t)&=\frac{1}{\hbar}\sum_k {\sigma_{x_k}\sigma_{p_k}}\omega_k^2 x_{0_k}^{(a,b)}\deltaxk{c}{d}\sin(\omega_k t)\\
\mathcal{J}^{ab}_{cd}(t)&=\sum_k \omega_k^2 x_{0_k}^{(a,b)}\xbark{c}{d}\cos(\omega_k t) \\
\mathcal{K}^{ab}_{cd}(t)&=\sum_k \omega_k^2 x_{0_k}^{(a,b)}\deltaxk{c}{d}\cos(\omega_k t) \\
\mathcal{L}^{ab}_{cd}(t)&=\sum_k \sigma_{x_k} \sigma_{p_k}\omega_k^3 x_{0_k}^{(a,b)}x_{0_k}^{(c,d)}\cos(\omega_k t)\\
\mathcal{M}^{ab}_{cd}(t)&=\hbar\sum_k \omega_k^3 x_{0_k}^{(a,b)}x_{0_k}^{(c,d)}\sin(\omega_k t),
\end{align}
where the generalization of these expressions to continuum environments is straightforwardly done by defining the necessary generalized spectral densities. With these definitions, we see that the eigenbasis Liouville pathway weights for first and second order in the perturbative expansion can be expressed as
\begin{equation}\label{eq:RedfieldLinearLiouvilleCF}
 \chi^{(1)}_{\{\alpha_j\alpha'_j\}}(\{\tau_j\}_{j=0}^{2}) = \mathcal{N}^{\alpha_1\alpha_0}_{1,1}(\{\tau_j\}_{j=0}^{2})\delta_{\alpha'_0\alpha'_1} - \mathcal{N}^{\alpha'_0\alpha'_1}_{1,1}(\{\tau_j\}_{j=0}^{2})\delta_{\alpha_1\alpha_0}
\end{equation}
and
\begin{equation}\label{eq:RedfieldSecondLiouvilleCF}
\begin{split}
\chi^{(2)}_{\{\alpha_j\alpha'_j\}}(\{\tau_j\}_{j=0}^{3}) = & \left( \mathcal{N}^{\alpha_2\alpha_1}_{2,2}(\{\tau_j\}_{j=0}^{3})\delta_{\alpha'_1\alpha'_2} - \mathcal{N}^{\alpha'_1\alpha'_2}_{2,2}(\{\tau_j\}_{j=0}^{3})\delta_{\alpha_2\alpha_1} \right) \\
& \times \left(  \mathcal{N}^{\alpha_1\alpha_0}_{1,2}(\{\tau_j\}_{j=0}^{3})\delta_{\alpha'_0\alpha'_1} - \mathcal{N}^{\alpha'_0\alpha'_1}_{1,2}(\{\tau_j\}_{j=0}^{3})\delta_{\alpha_1\alpha_0} \right)  + \mathcal{P}(\tau_2-\tau_1).
\end{split}
\end{equation}
Here, we have defined 
\begin{equation}\label{eq:NFunction}
\begin{split}
\mathcal{N}^{ab}_{j,N} = - \bigg(i\sum_{i=0}^N  & \left\{  \mathcal{I}^{ab}_{\alpha_i\alpha'_i}(\tau_{i+1}-\tau_j) -  \mathcal{I}^{ab}_{\alpha_i\alpha'_i}(\tau_{i}-\tau_j) \right\}  + \sum_{i=0}^{j-1} \left\{  \mathcal{J}^{ab}_{\alpha_i\alpha'_i}(\tau_{i+1}-\tau_{j}) -  \mathcal{J}^{ab}_{\alpha_i\alpha'_i}(\tau_{i}-\tau_j) \right\}\\
& - \frac{1}{2}S_j\sum_{i=j}^N   \left\{ \mathcal{K}^{ab}_{\alpha_i\alpha'_i}(\tau_{i+1}-\tau_{j}) -  \mathcal{K}^{ab}_{\alpha_i\alpha'_i}(\tau_{i}-\tau_j) \right\} \bigg)
\end{split}
\end{equation}
and
\begin{equation}\label{eq:PFunction}
\begin{split}
\mathcal{P}(\tau_2-\tau_1) & =  \left(\mathcal{L}^{\alpha_2\alpha_1}_{\alpha_1\alpha_0}(\tau_2-\tau_1)  - \frac{i}{2}S_1 \mathcal{M}^{\alpha_2\alpha_1}_{\alpha_1\alpha_0}(\tau_2-\tau_1) \right) \delta_{\alpha'_0\alpha'_1}\delta_{\alpha'_1\alpha'_2} \\
& - \left(\mathcal{L}^{\alpha_2\alpha_1}_{\alpha'_0\alpha'_1}(\tau_2-\tau_1)  - \frac{i}{2}S_1 \mathcal{M}^{\alpha_2\alpha_1}_{\alpha'_0\alpha'_1}(\tau_2-\tau_1) \right) \delta_{\alpha_1\alpha_0}\delta_{\alpha'_1\alpha'_2}\\
& - \left(\mathcal{L}^{\alpha_1\alpha_0}_{\alpha'_1\alpha'_2}(\tau_2-\tau_1)  - \frac{i}{2}S_1 \mathcal{M}^{\alpha_1\alpha_0}_{\alpha'_1\alpha'_2}(\tau_2-\tau_1) \right) \delta_{\alpha_2\alpha_1}\delta_{\alpha'_0\alpha'_1}\\
& +  \left(\mathcal{L}^{\alpha'_0\alpha'_1}_{\alpha'_1\alpha'_2}(\tau_2-\tau_1)  - \frac{i}{2}S_1 \mathcal{M}^{\alpha'_0\alpha'_1}_{\alpha'_1\alpha'_2}(\tau_2-\tau_1) \right) \delta_{\alpha_2\alpha_1}\delta_{\alpha_1\alpha_0}
\end{split}
\end{equation}
as Liouville pathway-dependent quantities that ensure the appropriate correlation functions contribute.
}
\section*{Acknowledgements}
This work was supported as part of the Center for Molecular Quantum Transduction, an Energy Frontier Research Center funded by the U.S. Department of Energy, Office of Science, Basic Energy Sciences, under Award No. DE-SC0021314.

\end{document}


\date{\today}
\flushbottom

\title{Supporting Information: Perturbation theory under the truncated Wigner approximation reveals how system-environment entanglement formation drives quantum decoherence} 
\author{Justin Provazza}
\affiliation{Department of Chemistry, Northwestern University, Evanston, Illinois 60208, USA}
\author{Roel Tempelaar}
\affiliation{Department of Chemistry, Northwestern University, Evanston, Illinois 60208, USA}
\maketitle
\noindent \textbf{This PDF includes}:
\begin{itemize}
\item[S1] Wigner analysis of the perturbative expansion and the emergence of Bopp operators from the path integral. 
\item[S2] Derivation of the Weyl symbol for harmonic oscillator projection operators.
\end{itemize}

\section{Wigner analysis of the perturbative expansion and the emergence of Bopp operators from the path integral:}
\label{sisec:expansion}
In order to obtain the expression of Eq. (6) of the main text, it is sufficient to show the result of the first order term in the expansion, as higher order terms follow trivially from this analysis. For conciseness, we simply consider the following contribution to the first order expectation value
\begin{equation}\label{si:propagator1}
\begin{split}
& \sum_{n_t}\Trenv\left\{\bra{n_t}\hat{O}e^{-\frac{i}{\hbar}\hat{H}_0(t-\tau_1)}\hat{V}(\tau_1)e^{-\frac{i}{\hbar}\hat{H}_0\tau_1}\hat{\rho}(0)e^{\frac{i}{\hbar}\hat{H}_0 t}\ket{n_t}\right\}\\
& =\sum_{n_t}\Trenv \bigg\{\sum_{\substack{n_0,n'_0 \\ n_1,n'_1}}\sum_{m_0,m'_0}  \bra{n_t} \hat{O}e^{-\frac{i}{\hbar}\hat{H}_0(t-\tau_1)}\ket{n_1}V_{n_1m_0}(\vec{\hat{x}},\tau_1)\bra{m_0}e^{-\frac{i}{\hbar}\hat{H}_0\tau_1}\ket{n_0} \\
& \hspace{3cm}\times \bra{n_0}\hat{\rho}(0)\ket{n'_0} \bra{n'_0}e^{\frac{i}{\hbar}\hat{H}_0\tau_1}\ket{m'_0}\braket{m'_0|n'_1}\bra{n'_1}e^{\frac{i}{\hbar}\hat{H}_0(t-\tau_1)}\ket{n_t}\bigg\}
\end{split}
\end{equation}
Given the definition of the reference Hamiltonian in Eq. (2) in the main text, we can simplify the above expression to
\begin{equation}\label{si:propagator2}
\begin{split}
\sum_{n_t}\sum_{\substack{n_0,n'_0 \\ n_1,n'_1}}\delta_{n'_0n'_1} \delta_{n'_1n_t} \Trenv\bigg\{ \bra{n_t}\hat{O}\ket{n_1}&e^{-\frac{i}{\hbar}(\hat{T}+U_{n_1}(\vec{\hat{x}}))(t-\tau_1)}V_{n_1n_0}(\vec{\hat{x}},\tau_1)e^{-\frac{i}{\hbar}(\hat{T}+U_{n_0}(\vec{\hat{x}}))\tau_1}\\
& \times \bra{n_0}\hat{\rho}(0)\ket{n'_0} e^{\frac{i}{\hbar}(\hat{T}+U_{n'_0}(\vec{\hat{x}}))\tau_1}e^{\frac{i}{\hbar}(\hat{T}+U_{n'_1}(\vec{\hat{x}}))(t-\tau_1)}\bigg\}
\end{split}
\end{equation}
where $\hat{T}=\sum_k\frac{\hat{p}_k^2}{2m_k}$ is the kinetic energy of the environment. The product of delta functions become (a part of) a Liouville space pathway weight that defines an element of the $\Theta$-tensor in Eq. (6) of the main text. To proceed, we focus on the portion of Eq. (\ref{si:propagator2}) that contains environmental operator dependence. For arbitrary potential forms, one may always rigorously approximate the evolution of the environmental degrees-of-freedom (DOFs) as being governed by classical equations of motion through the so-called truncated Wigner approximation (TWA, also known as the Linearized Semiclassical Initial Value Representation). This approximation can be easily formulated in both the cartesian phase space considered here as well as coherent state phase space, and we refer to previous works for the details of the analysis.\cite{polkovnikov2010phase,Shi2004,huo2011communication,AnalyticVibronicSpectra} Within this formulation, time-evolved expectation values take a classical-like form (\textit{i.e.}, convolution of a function with a probability distribution function on phase space). However, rather than the purely classical function and Boltzmann distribution, the TWA corresponds to convoluting the \textit{Weyl symbol} of an operator with a \textit{Wigner quasi-probability distribution}, as defined in Eq. (7) (and the subsequent text) in the main text. From the path integral formalism, this derivation is performed simply by writing the phase space path integrals directly in terms of \textit{mean} and \textit{difference} variables with the following substitution
\begin{equation}\label{si:MeanAndDifference}
\int \Dd x(\tau) \Dd x'(\tau) \Dd p(\tau) \Dd p'(\tau) \rightarrow \int \Dd \bar{x}(\tau) \Dd z(\tau) \Dd \bar{p}(\tau) \Dd y(\tau). 
\end{equation}
Here, $\bar{x}(\tau) = \frac{1}{2} (x(\tau)+ x'(\tau))$ and $ z(\tau) = x(\tau)-x'(\tau)$ where $x(\tau)$ ($x'(\tau)$) labels the ``forward'' (``backward'') path. There are analogous relations defining the mean and difference momentum variable, $\bar{p}(\tau)$ and $y(\tau)$, respectively in terms of $p(\tau)$ and $p'(\tau)$. Rewriting the action in terms of these mean and difference variables and expanding the difference of forward and backward actions to linear order in $z(\tau)$ about the mean path $\bar{x}(\tau)$ leads one to recover delta functional-constrained classical trajectories following analytic integration over all difference paths.

For harmonic potentials of the type considered here, one can treat the environmental difference path dependence of the perturbation operator analytically in an exact way within the momentum fluctuation framework introduced by Polkovnikov, \cite{polkovnikov2003quantum,polkovnikov2010phase} as previously applied for the calculation of vibronic spectral line shapes.\cite{AnalyticVibronicSpectra} In those applications, however, the momentum fluctuations accounted for corrections to the time-evolution of DOFs to treat terms higher order terms in the expansion of the forward-backward action difference. Here, where there are no higher order terms, we utilize this framework to account for (potentially arbitrary order) environmental operator dependent perturbations.

We start by expanding a perturbation matrix element as 
\begin{equation}\label{si:PerturbationTaylor}
V_{n_1n_0}({\bar{x}}(\tau) \pm \frac{{z}(\tau)}{2},\tau) = \sum_{m=0}^\infty  V^{(m)}_{n_1n_0}({\bar{x}}(\tau),\tau) \frac{(\pm z(\tau)/2)^m}{m!},
\end{equation}
where, for simplicity in notation, we have assumed a one-dimensional problem. The $\pm$ sign in the above expression arises as necessary from expressing the path integral in terms of mean and difference paths. As such, if the perturbation operator acts on the left the expansion adopts a ``$+$'' sign and when it acts on the right it adopts a ``$-$'' sign due to the definitions below Eq. (\ref{si:MeanAndDifference}). This sign dependence ultimately generates the causal relations for time-dependent Bopp operators used to evaluate non-equal time correlation functions.\cite{polkovnikov2010phase} To proceed we express the environment-dependent part of Eq. (\ref{si:propagator2}) in terms of a discrete time path integral in mean and difference variables, taking $\tau_1 = \tau_{N_1} = N_1\Delta\tau$ where $\Delta\tau=t/N$ is a discrete time element (with $0\le N_1\le N$), as
\begin{equation}\label{si:propagatoragain}
\begin{split}
&\Trenv\bigg\{ \hat{O}^{n_tn_1}e^{-\frac{i}{\hbar}(\hat{T}+U_{n_1}(\vec{\hat{x}}))(t-\tau_1)}V_{n_1n_0}(\vec{\hat{x}},\tau_1)e^{-\frac{i}{\hbar}(\hat{T}+U_{n_0}(\vec{\hat{x}}))\tau_1}\hat{\rho}_{\text{env}}^{n_0,n'_0} e^{\frac{i}{\hbar}(\hat{T}+U_{n'_0}(\vec{\hat{x}}))\tau_1}e^{\frac{i}{\hbar}(\hat{T}+U_{n'_1}(\vec{\hat{x}}))(t-\tau_1)}\bigg\} \\
& = \int d\bar{x}_0 \frac{d\bar{p}_0}{2\pi\hbar} \frac{dy_0}{2\pi\hbar}  d\bar{x}_N \left(\prod_{j=1}^{N-1}\int d\bar{x}_j\frac{d\bar{p}_j}{2\pi\hbar} dz_j\frac{dy_j}{2\pi\hbar}\right) O_W^{n_t,n_1}(\bar{x}_N,\bar{p}_{N-1}) W^{n_0,n'_0}({x}_0,{p}_0) \\
&\hspace{1cm} \times e^{-\frac{i}{\hbar}\sum_{j=1}^{N_1}\Delta\tau [U_{n_0}(\bar{x}_j) - U_{n'_0}(\bar{x}_j)]} e^{-\frac{i}{\hbar}\sum_{j=N_1+1}^{N}\Delta\tau [U_{n_1}(\bar{x}_j) - U_{n'_1}(\bar{x}_j)]}e^{\frac{i}{\hbar}\sum_{j=0}^{N-1}\Delta\tau\left[\left(\frac{\bar{x}_{j+1}-\bar{x}_{j}}{\Delta\tau}\right) - {\bar{p}_j}\right]y_j} \\ 
& \hspace{1cm} \times e^{-\frac{i}{\hbar}\sum_{j=1}^{N_1}\Delta\tau [(\frac{\bar{p}_{j}-\bar{p}_{j-1}}{\Delta\tau})-F^{(0)}_j]z_j} e^{-\frac{i}{\hbar}\sum_{j=N_1+1}^{N-1}\Delta\tau [(\frac{\bar{p}_{j}-\bar{p}_{j-1}}{\Delta\tau})-F^{(1)}_j]z_j} V_{n_1n_0}({\bar{x}_{N_1}} + \frac{{z}_{N_1}}{2},\tau_{N_1}).
\end{split}
\end{equation}
Here, $F^{(i)}_j = -\frac{1}{2}\frac{\partial}{\partial x_j}(U_{n_i}(x)+U_{n'_i}(x))|_{x=\bar{x}_j}$ is the active force during the $i$th time segment. One can immediately proceed by analytically performing integrals over difference variables other than the $j=N_1$ position difference variable, $\{z_j\}_{j\ne N_1}$ and $\{y_j\}_{j=0}^{N-1}$, resulting in delta functional constraints relating sequential mean position and momentum variables through the velocity and force, respectively, resulting in 
\begin{equation}\label{si:beforeidentity1}
\begin{split}
&\int d\bar{x}_0 \frac{d\bar{p}_0}{2\pi\hbar} d\bar{x}_N dz_{N_1}\left(\prod_{j=1}^{N-1}\int d\bar{x}_j\frac{d\bar{p}_j}{2\pi\hbar}\right) O_W^{n_t,n_1}(\bar{x}_N,\bar{p}_{N-1}) W^{n_0,n'_0}({x}_0,{p}_0)e^{-\frac{i}{\hbar}\sum_{j=1}^{N_1}\Delta\tau [U_{n_0}(\bar{x}_j) - U_{n'_0}(\bar{x}_j)]} \\
& \times  e^{-\frac{i}{\hbar}\sum_{j=N_1+1}^{N}\Delta\tau [U_{n_1}(\bar{x}_j) - U_{n'_1}(\bar{x}_j)]} \left(\prod_{j=1}^{N_1-1} \delta\left(\bar{p}_{j}-\bar{p}_{j-1}-\Delta\tau F^{(0)}_j\right) \right)\left(\prod_{j=N_1+1}^{N-1} \delta\left(\bar{p}_{j}-\bar{p}_{j-1}-\Delta\tau F^{(1)}_j\right) \right)\\
& \times \left( \prod_{j=0}^{N-1}\delta\left(\bar{x}_{j+1}-\bar{x}_{j}- \Delta \tau {\bar{p}_j}\right)\right) V_{n_1n_0}({\bar{x}_{N_1}} + \frac{{z}_{N_1}}{2},\tau_{N_1}) e^{-\frac{i}{\hbar}\Delta\tau [\bar{p}_{N_1}-\bar{p}_{N_1-1}-\Delta\tau F^{(0)}_{N_1}]z_{N_{1}}}.
\end{split}
\end{equation}
Next, we recognize that one can utilize the final phase factor in the above equation as a ``source'' term for generating $z_{N_1}$ terms by taking derivatives with respect to $\bar{p}_{N_1}$ so that, using the result of Eq. (\ref{si:PerturbationTaylor}), and finally performing the integral over $z_{N_1}$, we recover
\begin{equation}\label{si:beforeidentity2}
\begin{split}
&\int d\bar{x}_0 \frac{d\bar{p}_0}{2\pi\hbar} d\bar{x}_N \left(\prod_{j=1}^{N-1}\int d\bar{x}_j\frac{d\bar{p}_j}{2\pi\hbar}\right) O_W^{n_t,n_1}(\bar{x}_N,\bar{p}_{N-1}) W^{n_0,n'_0}({x}_0,{p}_0)e^{-\frac{i}{\hbar}\sum_{j=1}^{N_1}\Delta\tau [U_{n_0}(\bar{x}_j) - U_{n'_0}(\bar{x}_j)]} \\
& \times  e^{-\frac{i}{\hbar}\sum_{j=N_1+1}^{N}\Delta\tau [U_{n_1}(\bar{x}_j) - U_{n'_1}(\bar{x}_j)]} \left(\prod_{j=1}^{N_1-1} \delta\left(\bar{p}_{j}-\bar{p}_{j-1}-\Delta\tau F^{(0)}_j\right) \right)\left(\prod_{j=N_1+1}^{N-1} \delta\left(\bar{p}_{j}-\bar{p}_{j-1}-\Delta\tau F^{(1)}_j\right) \right)\\
&  \times \left( \prod_{j=0}^{N-1}\delta\left(\bar{x}_{j+1}-\bar{x}_{j}- \Delta \tau {\bar{p}_j}\right)\right) \left\{V_{n_1n_0}({\bar{x}_{N_1}} + \frac{i\hbar}{2}\frac{\partial}{\partial p_{N_1}},\tau_{N_1}) \delta\left(\bar{p}_{N_1}-\bar{p}_{N_1-1}-\Delta\tau F^{(0)}_{N_1}\right)\right\}
\end{split}
\end{equation}
Finally, with this result, we can continue by performing analytical integration over all mean variables other than $\bar{p}_{N_1}$, $\bar{p}_{0}$, and $\bar{x}_{0}$. By doing this, through the delta-constraints relating sequential position and momentum values, all terms containing dependence on mean momentum and position variables at times later than $\tau_{N_1}$ become functions of $\bar{p}_{N_1}$ (along with $\bar{p}_{0}$, and $\bar{x}_{0}$) and we can use the relation
\begin{equation}\label{si:deltaidentity}
\int dx f(x) \ \frac{\partial^n}{\partial x^n} \delta(x) = (-1)^n \int dx \ \delta(x) \left( \frac{\partial^n}{\partial x^n} f(x)\right), 
\end{equation}
which is simply a result of successive integration-by-parts. At this point we still can not generally perform the analytic integration over $\bar{p}_{N_1}$. However, we can alleviate this issue through the momentum fluctuation analysis\cite{polkovnikov2010phase} by noticing that
\begin{equation}\label{si:fluctuation}
\frac{\partial^nf(\bar{p}_{N_1} + \delta{p}_{N_1})}{\partial \delta{p}_{N_1}^n}\bigg|_{\delta{p}_{N_1}=0} = \frac{\partial^nf(\bar{p}_{N_1})}{\partial \bar{p}_{N_1}^n}.
\end{equation}
Thus, we introduce \textit{shifted} environmental trajectories, where the momentum shift occurs at the instance a perturbation operator acts on the quantum system. We take the $n$th order derivative of the function that depends on the shifted trajectory, and ultimately evaluate the resulting expression at $\delta{p}_{N_1}=0$.

Combining these two results and performing the mean momentum integration, we finally find
\begin{equation}\label{si:afteridentity}
\begin{split}
&\int d\bar{x}_0 \frac{d\bar{p}_0}{2\pi\hbar} \  W^{n_0,n'_0}({x}_0,{p}_0)  e^{-\frac{i}{\hbar}\int_0^{\tau_1}ds [U_{n_0}(x^{(0)}(s)) - U_{n'_0}(x^{(0)}(s))]}  \\
& \times  \bigg[V_{n_1n_0}(x^{(0)}(\tau_1) -\frac{i\hbar}{2}\frac{\partial}{\partial \delta p_{N_1}},\tau_{N_1}) O_W^{n_t,n_1}(x^{(1)}_{\delta p_{N_1}}(t),p^{(1)}_{\delta p_{N_1}}(t)) e^{-\frac{i}{\hbar}\int_{\tau_1}^{t}ds [U_{n_1}(x^{(1)}_{\delta p_{N_1}}(s)) - U_{n'_1}(x^{(1)}_{\delta p_{N_1}}(s))]}\bigg]_{\delta p_{N_1} = 0}.
\end{split}
\end{equation}
While the above result is general for arbitrary system state-dependent potentials, we can analytically express the shifted trajectories of harmonic environmental DOFs as
\begin{equation}\label{si:shiftedevolution}
x^{(i)}_{\delta p_k}(s) = x^{(i)}_{cl_k}(s) + \frac{1}{\omega_k}\sum_{j=1}^{i}\delta p^{(k)}_{\tau_j} \sin(\omega_k(s-\tau_j))
\end{equation}
in order to proceed with the analysis as outlined in the main text.

One can clearly recognize the emergence of Bopp operators in the argument of the perturbative operator in Eq. (\ref{si:afteridentity}), and if the perturbation had operated on the right, the derivative term would carry a positive sign. Hence, the general time-dependent Bopp operator can be obtained as $\hat{x}_k \rightarrow x^{(i)}_{\delta p_k}(\tau_{i+1}) - S_{i+1} \frac{i\hbar}{2}\frac{\partial}{\partial \delta p^{(k)}_{\tau_{i+1}}}$ where $S_{i+1}=+1$ if the operator acts on the left or $S_{i+1}=-1$ if it operates on the right. This analysis can also be expressed in terms of directional derivatives in order to remain consistent with time ordering. \cite{polkovnikov2010phase} We note that, if the perturbation operator were dependent on momentum operators, the analysis is straightforwardly extended in the same way with position fluctuations rather than momentum fluctuations.

It is easy to see how this result extends to higher order in the perturbative expansion. One simply evaluates all expressions along the shifted trajectory defined in Eq. (\ref{si:shiftedevolution}), replaces each perturbative operator with its Bopp operator-dependent analogue as 
\begin{equation}\label{si:perturbationBopp}
V_{nm}(\hat{x},\tau_{j}) = V_{nm}( x^{(j-1)}_{\delta p}(\tau_{j}) - \frac{i\hbar}{2}S_{j}\frac{\partial}{\partial \delta p_{\tau_j}},\tau_{j}),
\end{equation}
performs all of the necessary derivatives on all terms containing momentum fluctuation dependence, and finally evaluates the resulting expression at $\{\delta p_{\tau_j}\} = 0$.

\section{Derivation of the Weyl symbol for harmonic oscillator projection operators:}
\label{sisec:HarmonicRedmat}
Although the Weyl symbols for (diagonal) projection operators onto harmonic energy eigenstates are known to be of the form of Laguerre polynomials times a Gaussian, we are unaware of an existing closed-form expression for the Weyl symbol of an off-diagonal projection operator. These ``coherences'' are crucially important for constructing, for example, the reduced density matrix describing the time-evolving state of a harmonic environmental DOF. Here, we present a derivation of such an expression, while also highlighting the equivalence between the Bopp operator analysis and the standard Moyal product analysis for the deriving the expression.

For a harmonic system, the projection operator, $\ket{n}\bra{m}$, can be related to the ground state projection operator, $\ket{0}\bra{0}$, via successive operations with raising and lowering operators as
\begin{equation}\label{si:projection}
\ket{n}\bra{m}=\frac{1}{\sqrt{n!m!}}(\hat{a}^\dagger)^n \ket{0}\bra{0}(\hat{a})^m,
\end{equation}
which we partition as $\hat{A} \equiv (\hat{a}^\dagger)^n$ and $\hat{B} \equiv \ket{0}\bra{0}(\hat{a})^m$. The Weyl symbol of a product of operators is known to be given by the Moyal product as
\begin{equation}\label{si:moyalproduct}
(\hat{A}\hat{B})_W(a^*,a) = A_{W}(a^*,a) e^{\frac{1}{2}\hat{\Lambda}}  B_{W}(a^*,a),
\end{equation}
where we have introduced the so-called symplectic operator as $\hat{\Lambda} \equiv \frac{\ola\partial}{\partial a}\frac{\ora\partial}{\partial a^*}  - \frac{\ola\partial}{\partial a^*}\frac{\ora\partial}{\partial a}$ and we define the coherent state variable $a \equiv \sqrt{\frac{\omega}{2\hbar}}(x+i\frac{p}{\omega})$. In order to evaluate the Weyl symbol of the operator in Eq. (\ref{si:projection}), we must first find the Weyl symbol of the operator $\hat{B}$, as defined beneath Eq. (\ref{si:projection}), since it itself is a product of the ground state projection operator with $\hat{a}^m$. In the coherent state variables used here, this expression is found to be
\begin{equation}\label{si:BWeyl}
\begin{split}
{B}_W(a^*,a) &= 2 e^{-2a^*a} \left[\sum_{j=0}^{m}\frac{1}{j!} \left(\frac{-1}{2}\right)^j\left( \frac{\ola\partial}{\partial a^*}\frac{\ora\partial}{\partial a}\right)^j \right]a^m \\
& = 2 e^{-2a^*a} \left[\sum_{j=0}^{m}{m\choose j} \left(\frac{-1}{2}\frac{\ola\partial}{\partial a^*}\right)^j a^{m-j}\right] \\
& = 2 e^{-2a^*a} \left(a  - \frac{1}{2}\frac{\ola\partial}{\partial a^*}\right)^m \\
& =  2^{m+1}e^{-2a^*a}a^m
\end{split}
\end{equation}
where we have used the fact that only one of the contributions to the symplectic operator contribute (and only up the $m$th order) due to the derivative operations acting on the lowering operator. Between the second and third line we recognize the binomial coefficients that allow us to relate this analysis to the Bopp operators in the coherent state phase space representation (where we have used the directional derivatives described in the preceding section), and in evaluating the final expression we note that $\sum_{j=0}^{m}{m\choose j}=2^m$.

Next, we take the total Weyl symbol as described in Eq. (\ref{si:moyalproduct}), which proceeds similarly to the previous analysis, giving the final result as
\begin{equation}\label{si:finalexpressionweyl}
\begin{split}
\left(\ket{n}\bra{m}\right)_W(x,p)&= \frac{2^{m+1}}{\sqrt{n!m!}} \left[\sum_{i=0}^{n}{n\choose i}(a^*)^{n-i} \left(\frac{-1}{2}\frac{\ora\partial}{\partial a}\right)^i \right] e^{-2a^*a}a^m \\
&=  \frac{2^{m+1}}{\sqrt{n!m!}} \sum_{i=0}^{n}\sum_{k=0}^{i}{n\choose i} {i\choose k} \left(\frac{-1}{2}\right)^i  (a^*)^{n-i} \left[\left(\frac{\ora\partial}{\partial a}\right)^k a^m\right] \left[\left(\frac{\ora\partial}{\partial a}\right)^{i-k}e^{-2a^*a}\right] \\
& = \frac{2^{m+1}}{\sqrt{n!m!}} \sum_{i=0}^{n}\sum_{k=0}^{\min\{i,m\}} k! {n\choose i} {i\choose k} {m \choose k}\left(\frac{-1}{2}\right)^k  (a^*)^{n-k} a^{m-k}e^{-2a^*a}.
\end{split}
\end{equation}
For diagonal elements, $n=m$, the expression above reduces to those known to be given by Laguerre polynomials while for off-diagonal elements, $n\ne m$, the expression describes coherences.

\bibliography{../bibliography}